\documentclass[preprint,12pt]{elsarticle}
\usepackage{}

\usepackage{amsmath}
\usepackage{cases}
\usepackage{color}
\usepackage{hyperref}
\usepackage{amssymb}
\usepackage{graphicx}
\usepackage{graphics}
\usepackage{subfigure}
\usepackage{slashbox}
\usepackage{appendix}
\usepackage{textcomp}
\usepackage{bm}
\usepackage{url}

\makeatletter
  \newcommand\figcaption{\def\@captype{figure}\caption}
  \newcommand\tabcaption{\def\@captype{table}\caption}
\makeatother

\biboptions{numbers,sort&compress}

\begin{document}

\begin{center}
{\bf \large Ultra-chaos: an insurmountable objective obstacle  of reproducibility and replicability} \\ \vspace{0.5cm}

Shijun Liao$^{1,2,3}$ \footnote{The corresponding author,  email address: sjliao@sjtu.edu.cn} and Shijie Qin $^{2}$  \\ \vspace{0.3cm}
$^{1}$ State Key Laboratory of Ocean Engineering, Shanghai 200240, China\\
$^{2}$ Center of Marine Numerical Experiment, School of Naval Architecture, Ocean and Civil Engineering, Shanghai Jiaotong University,  China\\
$^{3}$ School of Physics and Astronomy,  Shanghai Jiaotong University,  China 
\end{center}

\hspace{-0.75cm} {\bf Abstract}
{\em In this paper, a new concept, i.e. ultra-chaos, is proposed for the first time.  Unlike a normal-chaos,  statistical properties such as the probability density functions (PDF) of an ultra-chaos are sensitive to tiny disturbances.  We illustrate that ultra-chaos is widely existed and thus has general scientific meanings.  It is found that statistical non-reproducibility is an inherent property of an ultra-chaos so that an ultra-chaos is  at a higher-level of disorder than a normal-chaos.   Thus, it is impossible in  practice to replicate experimental/numerical results of an ultra-chaos even in statistical meanings, since random environmental noises always exist and are out of control.  Thus,  the ultra-chaos should be an insurmountable obstacle of reproducibility and replicability.   Similar to G\"{o}del's incompleteness theorem, such kind of ``incompleteness of reproducibility'' reveals a limitation of our traditional scientific paradigm based on reproducible experiments, which can be traced back to Galileo.  The ultra-chaos opens a new door and possibility to study chaos theory, turbulence theory,  computational fluid dynamics (CFD),  the statistical significance, reproducibility crisis,  and so on. }\\ \vspace{0.1cm}

\hspace{-0.75cm}{\bf Keywords}  {Chaos, statistical stability, reproducibility and replication.}\\ \vspace{0.1cm}

\hspace{-0.75cm}{\bf AMS}  {35N05, 65P20}

\section{Introduction}

The chaos theory \cite{Poincare1890, Lorenz1963, Li1975Period, Rossler1979, Eckmann1985RMP, Lorenz1993Book, PeterSmith1998Explaining, Sprott2003Chaos, sprott2010, Christophe2021Chaos} is widely regarded as the third greatest scientific revolution in physics in 20th century, comparable to Einstein's theory of relativity  and the quantum mechanics.   It is Poincar{\' e} \cite{Poincare1890} who first discovered  ``the sensitivity dependence on initial conditions'' (SDIC) of chaotic systems, which was rediscovered by Lorenz \cite{Lorenz1963}  with a popular name ``butterfly-effect'':  the exact time of formation and  the exact path of a tornado might be influenced by tiny disturbances  such as a distant butterfly that flapped its wings several weeks earlier.  In addition, Lorenz \cite{Lorenz2006} further discovered ``the sensitivity dependence on numerical algorithms'' (SDNA) of chaotic systems: computer-generated chaotic numerical simulations given by different algorithms in single/double precision quickly depart from each other with distinct deviations.  Naturally, such kind of non-replicability of chaotic trajectory  leaded to some heated debates on the credence of numerical simulations of chaos, and brought a crisis of confidence: some even made a rather pessimistic conclusion that ``for chaotic systems, numerical convergence cannot be guaranteed {\em forever}'' \cite{Teixeira2007}.

In order to gain a reproducible/reliable numerical simulations of chaotic trajectory,   Liao \cite{Liao2009} suggested  a numerical strategy,  namely  the ``Clean Numerical Simulation'' (CNS).   In the frame of the CNS \cite{Liao2009, Liao2013A, Liao2013B, LIAO2014On, li2014stability, Lin2017On,  li2017more, li2018over,  li2019collisionless, Hu2020JCP, Qin2020CSF, Xu2021PoF, li2021one, Liao2021arXiv},  the temporal/spatial truncation errors are  reduced to a required tiny level by means of a high {\em enough} order of Taylor expansion in time and a fine {\em enough} spatial discretization with spatial Fourier expansion, respectively.  Besides,  the round-off error is reduced  to a required tiny level by means of reserving a large {\em enough} number of significant digits for  all physical/numerical  variables and parameters in multiple precision \cite{oyanarte1990mp}.   Furthermore,  an additional simulation with even smaller numerical noises is needed to determine the so-called  ``critical predictable time''  $T_{c}$ by comparing these two simulations, so that the numerical noise is negligible and thus the computer-generated result is reproducible/reliable within the whole spatial domain in  the time interval $t\in[0,T_c]$.

\begin{figure}[tb]
  \centering
  \includegraphics[width=2.5in]{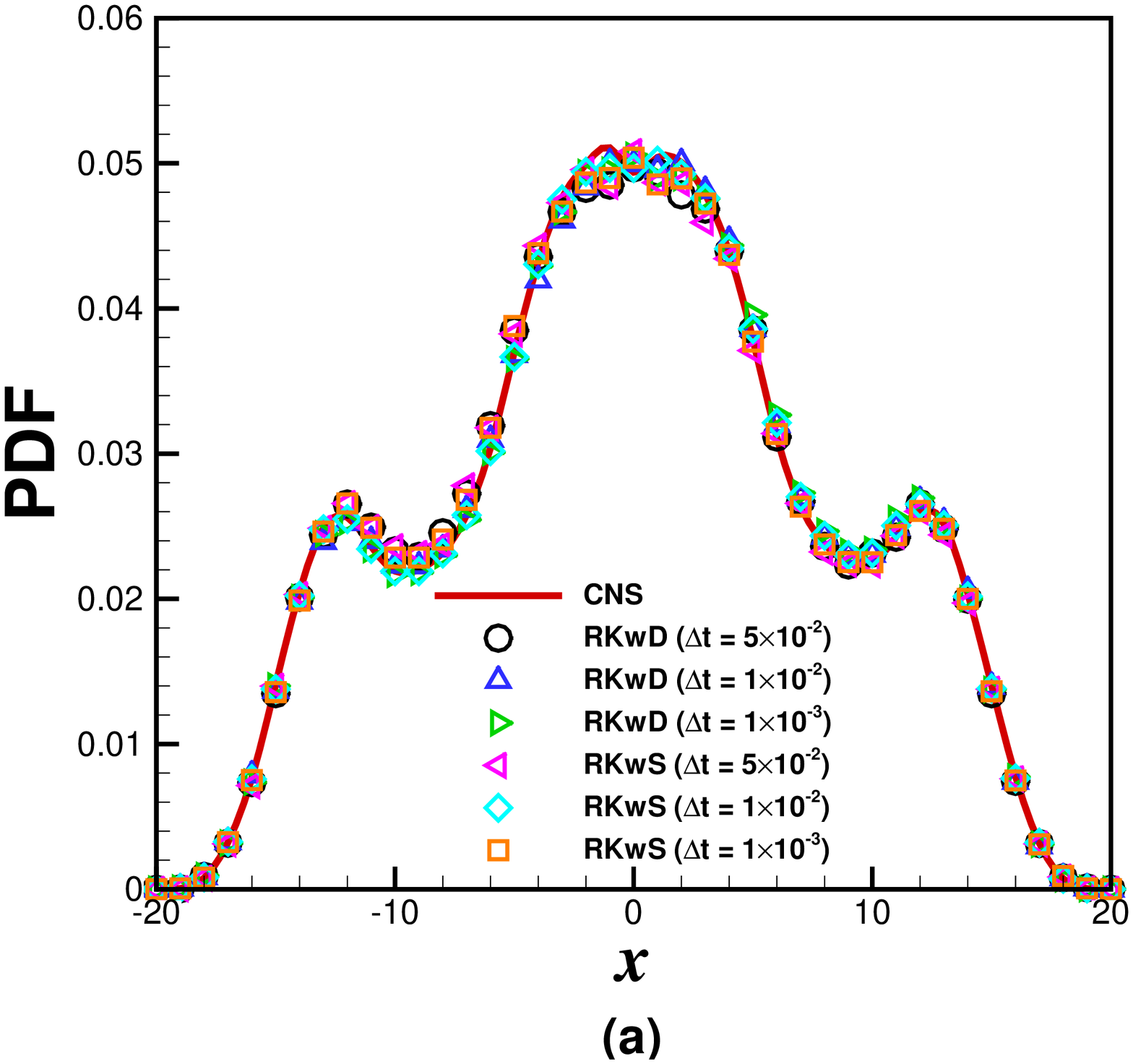}
  \includegraphics[width=2.5in]{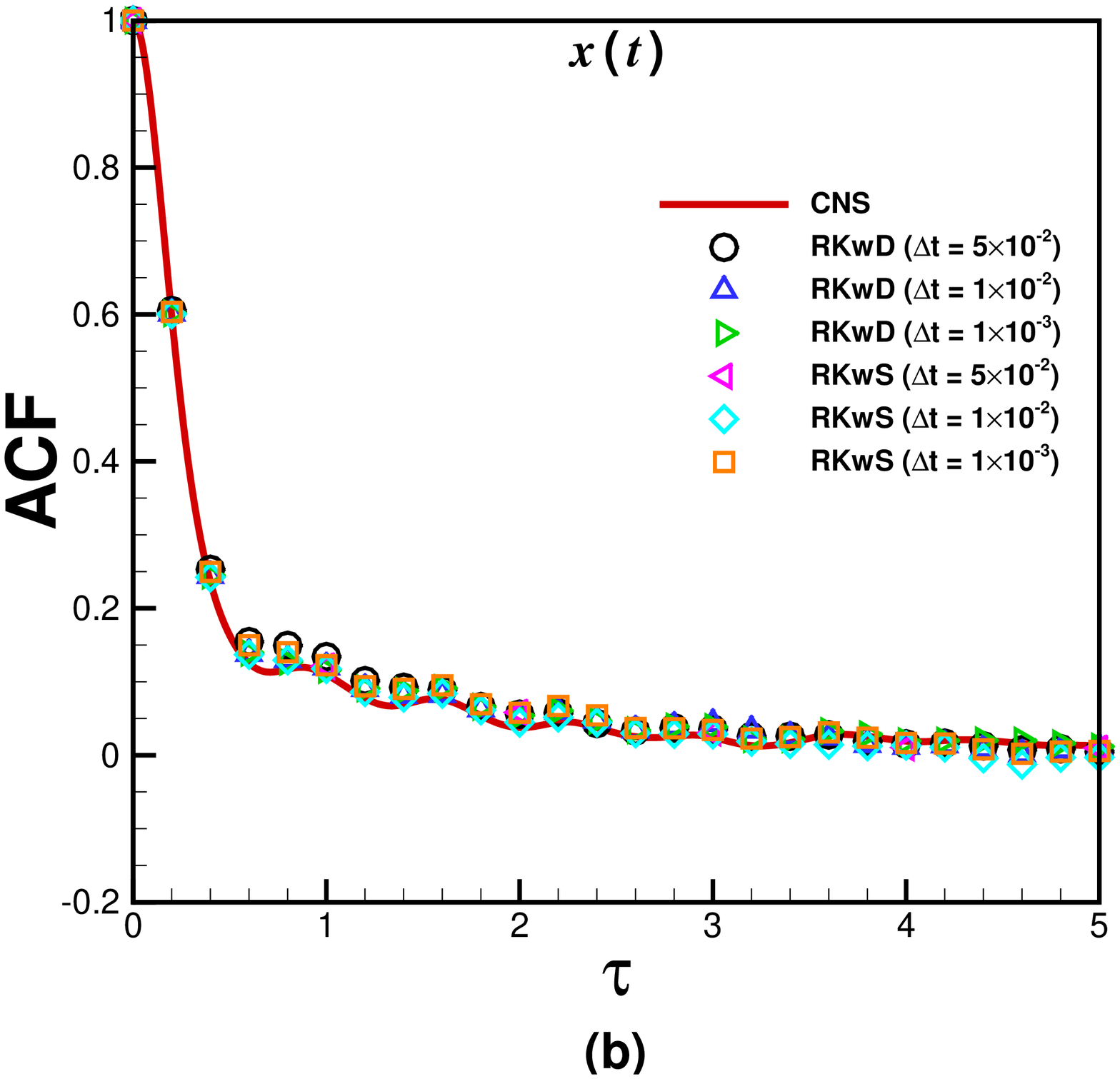}
  \caption{{\bf Influence of tiny noises to the statistics of a normal-chaos.}  (a) The probability density function (PDF) of $x(t)$; (b) The autocorrelation functions (ACF) of $x(t)$. The statistic results are based on the chaotic simulations $x(t)$ in $0 \leq t \leq 10000$  governed by the Lorenz equations (\ref{lorenz_x})-(\ref{lorenz_ini})  (with one positive Lyapunov exponents),  given by the CNS (red line) and the Runge-Kutta algorithms (symbols) with double-precision (RKwD) or single-precision (RKwS) using different time-step $\Delta t$.  }
  \label{fig-1}
\end{figure}

\begin{figure}[tb]
  \centering
  \includegraphics[width=2.5in]{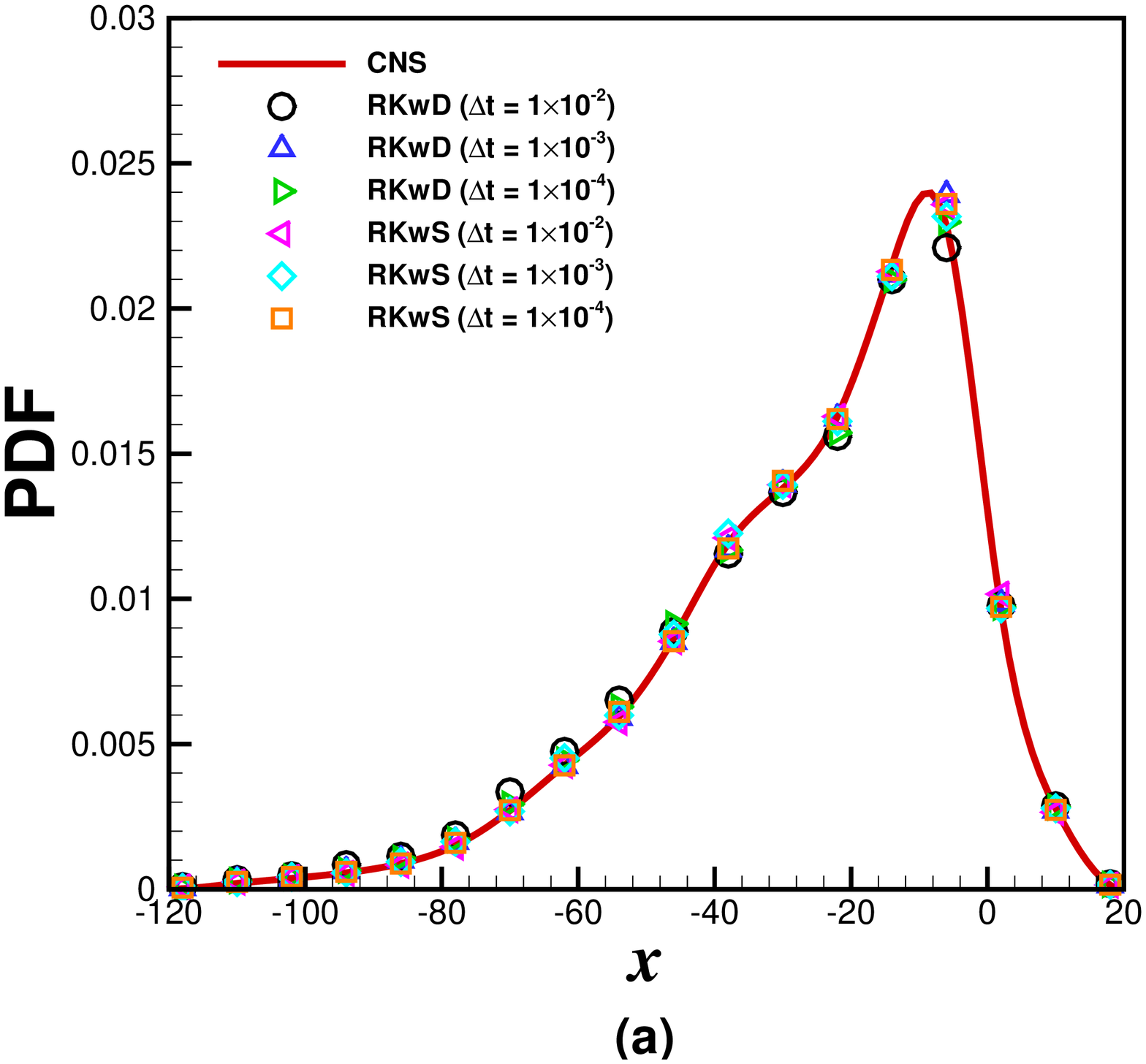}
  \includegraphics[width=2.5in]{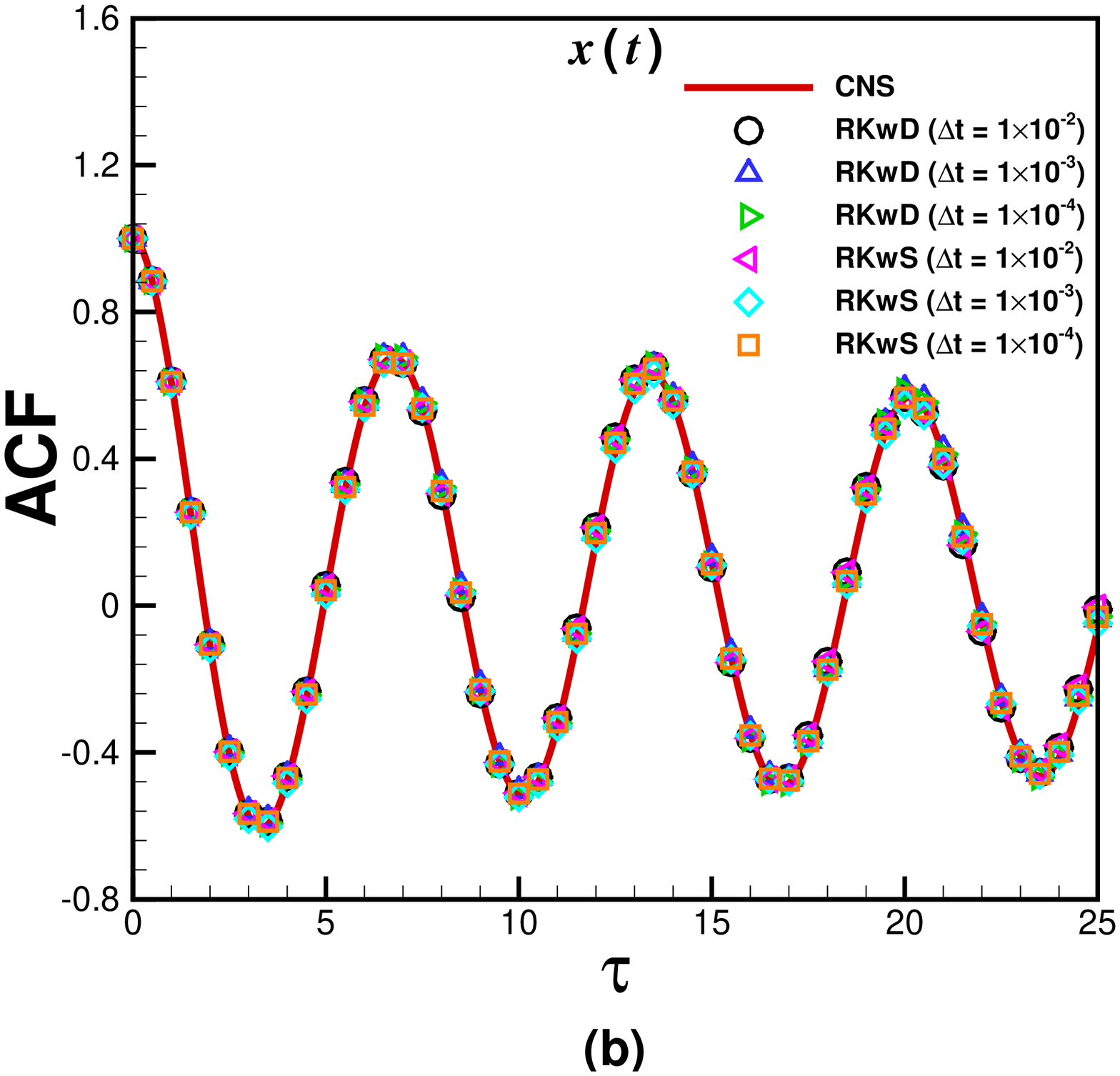}
  \caption{{\bf Influence of tiny noises to the statistics of a hyper-chaos.}    (a) The probability density function (PDF) of $x(t)$; (b) The autocorrelation functions (ACF) of $x(t)$.  The statistic results are based on the  simulations $x(t)$ in $0 \leq t \leq 10000$ governed by the four-dimensional R\"{o}ssler system (\ref{rossler_x})-(\ref{rossler_ini}) (with two positive Lyapunov exponents),  given by the CNS (red line) and the Runge-Kutta algorithms (symbols) with double-precision (RKwD) or single-precision (RKwS) using different time-step $\Delta t$.  }
  \label{fig-2}
\end{figure}

The CNS has been successfully applied to  gain reproducible/reliable simulations of  trajectories  of  many chaotic systems, such as Lorenz equation \cite{LIAO2014On},  Rayleigh-B{\'e}nard turbulent flows \cite{Lin2017On},  and some spatiotemporal chaotic systems related to  the complex Ginzburg-Landau equation \cite{Hu2020JCP},  the damped driven sine-Gordon equation \cite{Qin2020CSF} and the chaotic motion of a free fall  desk \cite{Xu2021PoF}.  Especially,  more than 2000 new periodic orbits of three-body system have been found by means of the CNS \cite{li2017more,li2018over,li2019collisionless}, which were reported twice by the popular magazine New Scientist \cite{NewScientist2017,NewScientist2018},  because only three families of periodic orbits of the three-body problem have been reported in three hundred years since Newton mentioned the famous problem in 1687.  All of these illustrate the validity of the CNS for chaos.

It should be emphasized that such a reproducible/reliable result  given by the CNS  provides us a ``true''  solution, and more importantly,  a  kind of  {\em reproducibility} and {\em replicability} of chaotic trajectory.    Such kind of  trajectory reproducibility and replicability can remain in a long enough interval  of time $t\in[0,T_c]$  as long as the background numerical noises could be globally reduced to a rather tiny level \cite{Hu2020JCP,Qin2020CSF,Xu2021PoF}.   This  kind of  strict  reproducibility and replicability of chaotic trajectory provides us a confidence of credence/reliability and especially a {\em benchmark} solution.  For example, by means of an algorithm based on the CNS, Liao and Wang \cite{LIAO2014On} gained a computer-generated chaotic simulation of Lorenz equation, which is reproducible/reliable in a quite long interval of time $0 \leq t  \leq 10000$ (Lyapunov  unit) and thus provides us a {\em true} result and a {\em benchmark} solution of the Lorenz equation in $t\in[0,10000]$, while  numerical simulations given by other algorithms in single/double precision  are  reproducible/reliable only in a quite smaller interval $0\leq t \leq 32$ (Lyapunov  unit) and has a distinct deviation from the benchmark solution for $t> 32$.  Having the benchmark solution of the Lorenz equation in such a long {\em enough} interval of time $t\in[0,10000]$,  it is possible to investigate the influence of numerical noise to {\em statistics} of computer-generated chaotic simulations given by other algorithms in single/double precision.

Note that, it has been widely believed (or assumed) that statistics of a chaos should be ``stable'' even if its trajectory is sensitive to small disturbances.  Mostly, this is indeed true, such as the chaos governed by the Lorenz equations (\ref{lorenz_x})-(\ref{lorenz_ini}) \cite{Liao2009, LIAO2014On} with one positive Lyapunov exponent, as shown in Figure~\ref{fig-1}, and the so-called {\em hyper-chaos} governed by the four-dimensional R\"{o}ssler system (\ref{rossler_x})-(\ref{rossler_ini}) with {\em two} positive Lyapunov exponents \cite{Stankevich2020Chaos}, as shown in Figure~\ref{fig-1}, respectively.  Note that Figure~\ref{fig-1} is given by Lorenz equations (with one positive Lyapunov exponents) 
\begin{align}
&  \dot{x}=\sigma\,(y-x),    \label{lorenz_x}    \\
&  \dot{y}=R\,x-y-x\,z,    \label{lorenz_y}    \\
&  \dot{z}=x\,y+B\,z    \label{lorenz_z}
\end{align}
in the case of $\sigma=10$, $R=28$, and $B=-8/3$, with the initial condition
\begin{equation}
x(0)=-15.8, \;\; y(0)=-17.48, \;\;  z(0)=35.64.    \label{lorenz_ini}
\end{equation}
Liao \& Wang \cite{LIAO2014On} gained a trajectory-replicable chaotic simulation of the above-mentioned  Lorenz equations in a quite long interval $ t \in [0,10000]$ (Lorenz unit time)  by means of a parallel algorithm of the CNS using the $3500$th-order Taylor expansion and the $4180$-digit multiple-precision for all physical/numerical parameters and variables, which is used here as the benchmark solution (marked by CNS).  For comparison,  we also solve the Lorenz equations (\ref{lorenz_x})-(\ref{lorenz_ini}) by means of the $4$th-order Runge-Kutta's method with double-precision (RKwD) or single-precision (RKwS) using different time-step $\Delta t$, corresponding to different levels of numerical/environmental noises.  

Figure~\ref{fig-2} is given by  the hyper-chaotic Rossler system (with two positive Lyapunov exponents) \cite{Stankevich2020Chaos}
\begin{align}
&  \dot{x}=-\,y-z,    \label{rossler_x}    \\
&  \dot{y}=x+a\,y+w,    \label{rossler_y}    \\
&  \dot{z}=b+x\,z,    \label{rossler_z}    \\
&  \dot{w}=-\,c\,z+d\,w    \label{rossler_w}
\end{align}
in the case of $a=0.25$, $b=3$, $c=0.5$ and $d=0.05$, with the initial condition
\begin{equation}
x(0)=-20, \;\; y(0)=z(0)=0, \;\;  w(0)=15.      \label{rossler_ini}
\end{equation}
We gained a trajectory-replicable chaotic simulation of this hyper-chaotic Rossler system  in a quite long interval  $t\in[0,10000]$ by means of a parallel algorithm of the CNS with the $200$th-order Taylor expansion and $500$-digit multiple-precision for all physical/numerical parameters and variables using the time-step $\Delta t = 0.001$, which is used here as the benchmark solution (marked by CNS).  For comparison,  we also solve this hyper-chaotic Rossler system by means of the $4$th-order Runge-Kutta's method with double-precision (RKwD) or single-precision (RKwS) using different time-step $\Delta t$, corresponding to different levels of numerical/environmental noises. 

It should be emphasized that, although the chaotic trajectories of the above-mentioned two chaotic systems are quite sensitive to small disturbances, their PDFs are {\em not} sensitive to numerical noises.    However,  it has been reported \cite{Hu2020JCP,Qin2020CSF}  that  even statistics  (such as  probability density function)  of  some  chaotic  systems  are  sensitive  to  numerical  noises.     
This  is  quite  unusual, which reveals the existence of chaotic systems that might be at a higher level of disorder.   We thought hard about this surprising result for a long time and realized that this kind of phenomena are so important that some new concepts should be proposed, since a new concept might be more fundamental than lots of examples.          
Thus, we define here the following new concepts: 
\begin{itemize}
\item normal-chaos: a chaotic system, whose statistics  are not sensitive to small disturbances.
\item ultra-chaos:   a chaotic system, whose  statistics  are sensitive to small disturbances.
\end{itemize}

Note that Figure~\ref{fig-2} clearly illustrates that the well-known ``hyper-chaos''  is just a normal-chaos, i.e. its statistics is not sensitive to the numerical/environmental noises.  This reveals the novelty of ultra-chaos, a new concept we proposed in this paper.   

In this paper, using the CNS as a powerful tool to investigate the influence of tiny external disturbances, we illustrate that the so-called ultra-chaos indeed widely exist.   This  fact  should have important  scientific meanings, as discussed in \S~3.       

\section{An illustrative example of ultra-chaos}

Note that, by means of the CNS,  the {\em artificial} background numerical noises can be globally reduced to a required tiny level much smaller than the {\em objective} environmental noises, so that we can accurately study the influence of  the environmental disturbances to the {\em statistics}  of some  chaos.   For example, let us consider here the damped driven sine-Gordon equation \cite{chacon2008spatiotemporal,Keller1995Surveys,ferre2017localized}, but  with a Gaussian white noise $\epsilon(x,t)$:
\begin{equation}
  u_{tt} = u_{xx} - \sin(u)  -\alpha u_{t} +\Gamma \sin (\omega \, t - \lambda \, x) + \epsilon(x,t)  \label{SGE}
\end{equation}
subject to a periodic boundary condition
\begin{equation}
  u(x+l, t)=u(x,t),   \label{periodic}
\end{equation}
where  the subscript denotes the derivative,  $x$ and $t$ are the spatial  and temporal variables,  $\alpha$ is a constant  related to the damped friction,  $\Gamma$ is a constant  related to the external force,   $\omega$ is the temporal frequency,  $\lambda = 2\pi/l$ is the spatial frequency with  $l$ being the total length of the system,  respectively.  Here,  $\epsilon(x,t)$ corresponds to the tiny environmental noises.  Without loss of generality, we  follow Chac\'{o}n {\em et al.} \cite{chacon2008spatiotemporal}  to choose the following values of these physical parameters
\begin{equation}
  \omega = \frac{3}{5}, \;  \alpha =\frac{1}{10}, \;\Gamma =\frac{461}{500}, \; l = 500, \; \lambda = \frac{2\pi}{l},  \label{parameters}
\end{equation}
subject to the zero initial condition
\begin{equation}
u(x,0)=0, \hspace{0.5cm} u_{t}(x, 0)=0.    \label{ini}
\end{equation}
As reported by Qin \& Liao \cite{Qin2020CSF}, when there are no environmental noises, i.e. $\epsilon(x,t) = 0$,
the above model corresponds to a spatiotemporal chaos, whose statistics are sensitive to numerical noises, and thus is an {\em ultra-chaos}. However, numerical noises are {\em artificial}  and  thus have no physical meanings.  So, in this paper, using the CNS as a powerful tool, we investigate the influence of tiny external disturbances to its statistical properties.  This has important meanings in physics, since tiny external disturbances are {\em unavoidable} in practice.  Let $\sigma_n$ denote its standard deviation of external disturbances $\epsilon(x,t)$ in Gaussian white noise.  We use the same CNS  algorithm \cite{Qin2020CSF} to gain reproducible/reliable  numerical simulations in the three cases $\sigma_n = 0$, $\sigma_n = 10^{-18}$ and $\sigma_n=10^{-20}$, respectively, where $\sigma_n =0$ corresponds to the benchmark solution given by the CNS without any objective environmental noise, the others correspond to the reproducible/reliable simulations with the influence of the corresponding  tiny environmental noises.   Unlike  Qin \& Liao \cite{Qin2020CSF} who used a CNS algorithm with $N = 16384$ and $N_{s} = 60$ to gain reliable simulations in $t\in[0,900]$,  we discretized the spatial domain here much better  by $N=2^{16}=65536$ equidistant points, used the multiple precision with $N_s = 230$ significant digits for all variables and parameters, and applied the variable stepsize scheme in the temporal dimension with a given allowed tolerance $tol = 10^{-230}$ of the governing equations.   In this way, the corresponding background numerical noises (related to truncation error and round-off error) are much less than the tiny external disturbances.  Besides, according to Qin \& Liao \cite{Qin2020CSF},  it holds
\[  T_c \approx \min \left\{ 0.0558 N +10.7, 16.5 N_s -87.8  \right\}.   \]
Thus, we can gain the reproducible/reliable  chaotic simulations  within the whole spatial domain in a much larger temporal interval $t\in[0,3600]$  in the above-mentioned three cases, which is long {\em enough} for our statistical analysis reported in this paper.

\begin{figure}[tb]
  \centering
  \includegraphics[width=2.5in]{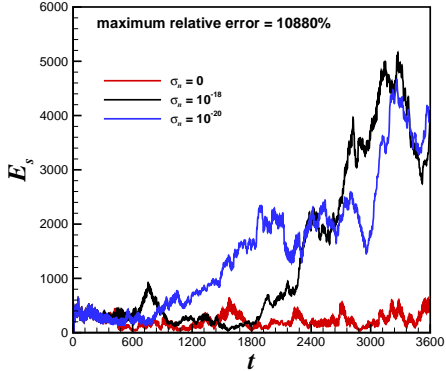}
  \caption{{\bf Influence of small disturbances to the total spectrum energy $E_{s}(t)$ of an ultra-chaos.}  The  curves are  based on the CNS results in $t\in[0,3600]$ of an ultra-chaos governed by the damped driven sine-Gordon equations (\ref{SGE})-(\ref{ini}) in case of  $\sigma_n=0$ (red), $\sigma_n=10^{-18}$ (black) and $\sigma_n=10^{-20}$ (blue), respectively, with the maximum relative error 10880\%, where $\sigma_{n}$ denotes the standard deviation of Gaussian white noise.}
  \label{fig-3}
\end{figure}

\begin{figure}[t]
  \centering
  \includegraphics[width=2.5in]{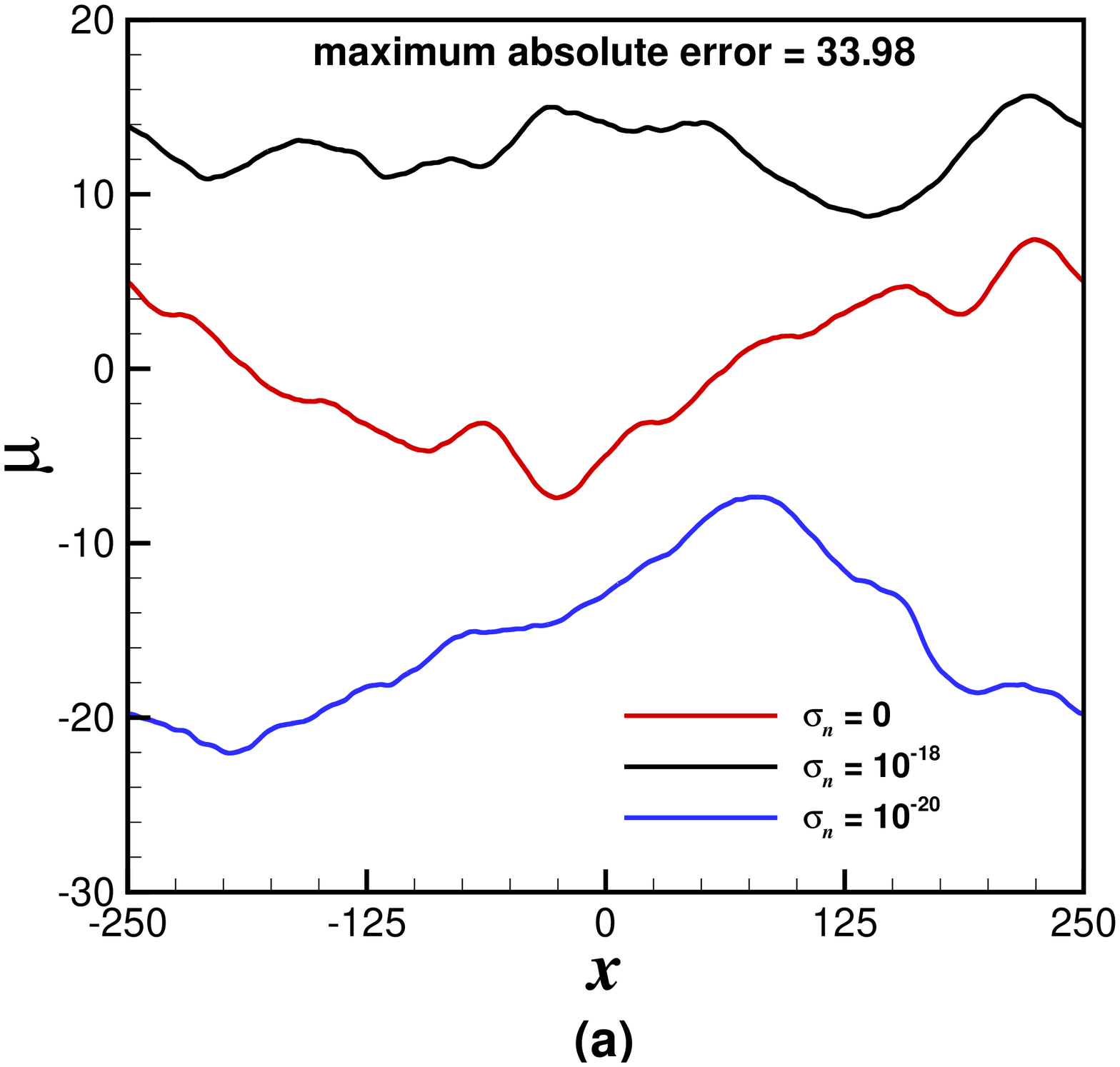}
  \includegraphics[width=2.5in]{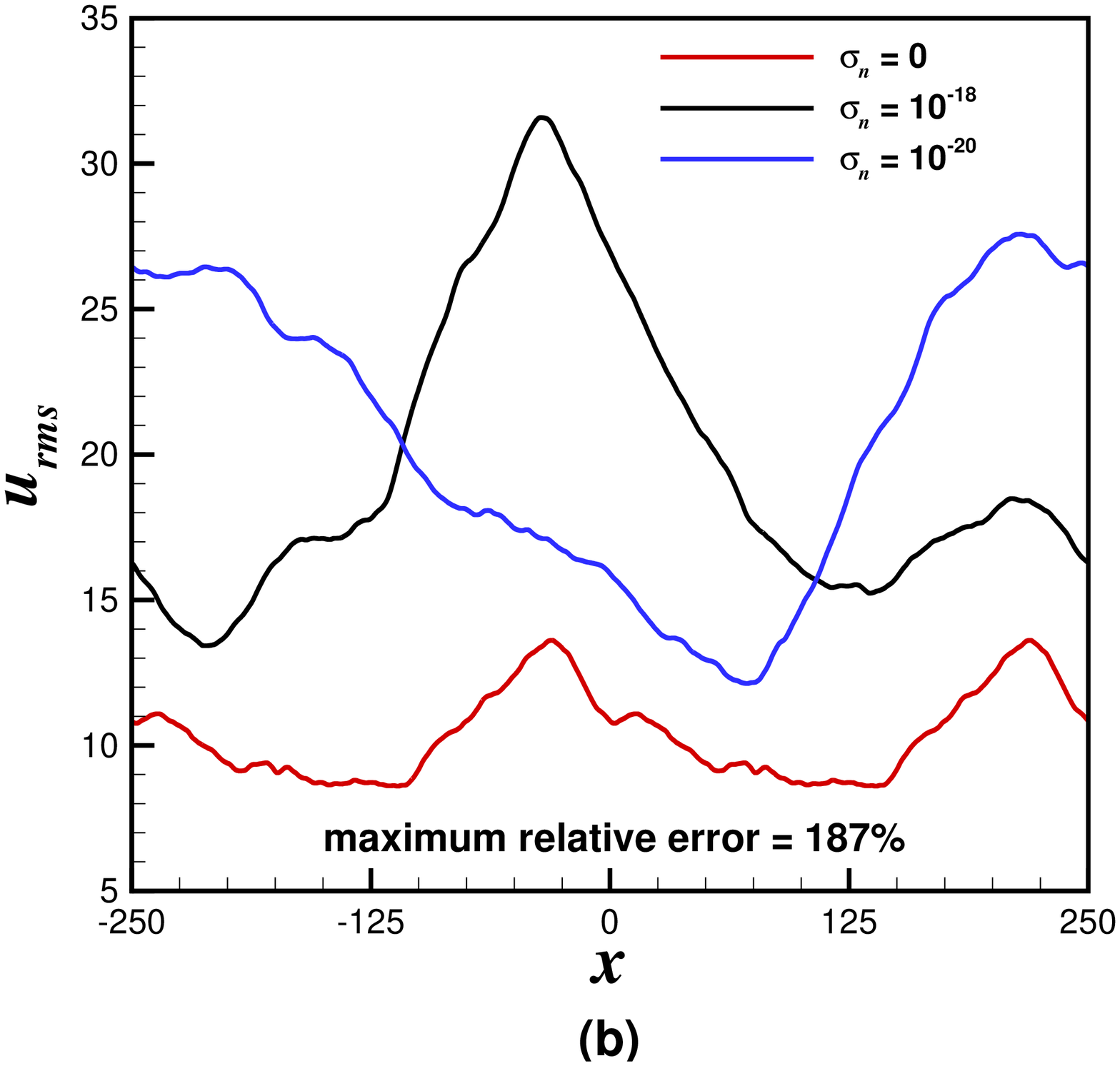}
  \caption{{\bf Influence of small disturbances to the temporal statistics  of an ultra-chaos.}  (a) The temporal mean $\mu(x)$; (b) The temporal Root-Mean-Square (RMS) $u_{rms}(x)$. The  curves are  based on the CNS results   in $t\in[0,3600]$ of an ultra-chaos governed by the damped driven sine-Gordon equations (\ref{SGE})-(\ref{ini}) in case of $\sigma_n=0$ (red), $\sigma_n=10^{-18}$ (black) and $\sigma_n=10^{-20}$ (blue), with the maximum absolute error $33.98$ for $\mu(x)$ and the maximum relative error $187\%$ for $u_{rms}(x)$, respectively, where $\sigma_{n}$ denotes the standard deviation of Gaussian white noise. }
  \label{fig-5}
\end{figure}

 Note that   $\sigma_n=0$ corresponds to no external disturbance.  Using the reproducible/reliable  result in case of $\sigma_n = 0$ as the benchmark solution, we can accurately investigate the influence of the tiny environmental noises in cases of $\sigma_n=10^{-18}$ and $\sigma_n=10^{-20}$, respectively.    It is found that the tiny environmental noises are equivalent to the numerical noises in essence: the small disturbances  indeed lead to huge and distinct  deviations of  chaotic trajectories  and   many  physical quantities from the benchmark solution (in case of $\sigma_n = 0$) in $t > 360$ when the tiny external disturbances are enlarged up to the macroscopic level.  
For example,  the total spectrum energy $E_{s}(t)$ in case of $\sigma_n=10^{-18}$ and $\sigma_n=10^{-20}$ have huge deviations from that of the benchmark solution (in  case of $\sigma_n=0$) with the maximum relative error 10880\%, as shown in Figure~\ref{fig-5}.  It indicates that, due to the existence of the tiny environmental noises, the ultra-chaotic system under consideration might contain much larger energy than that without any environmental noises.

\begin{figure}[tb]
  \centering
  \includegraphics[width=2.5in]{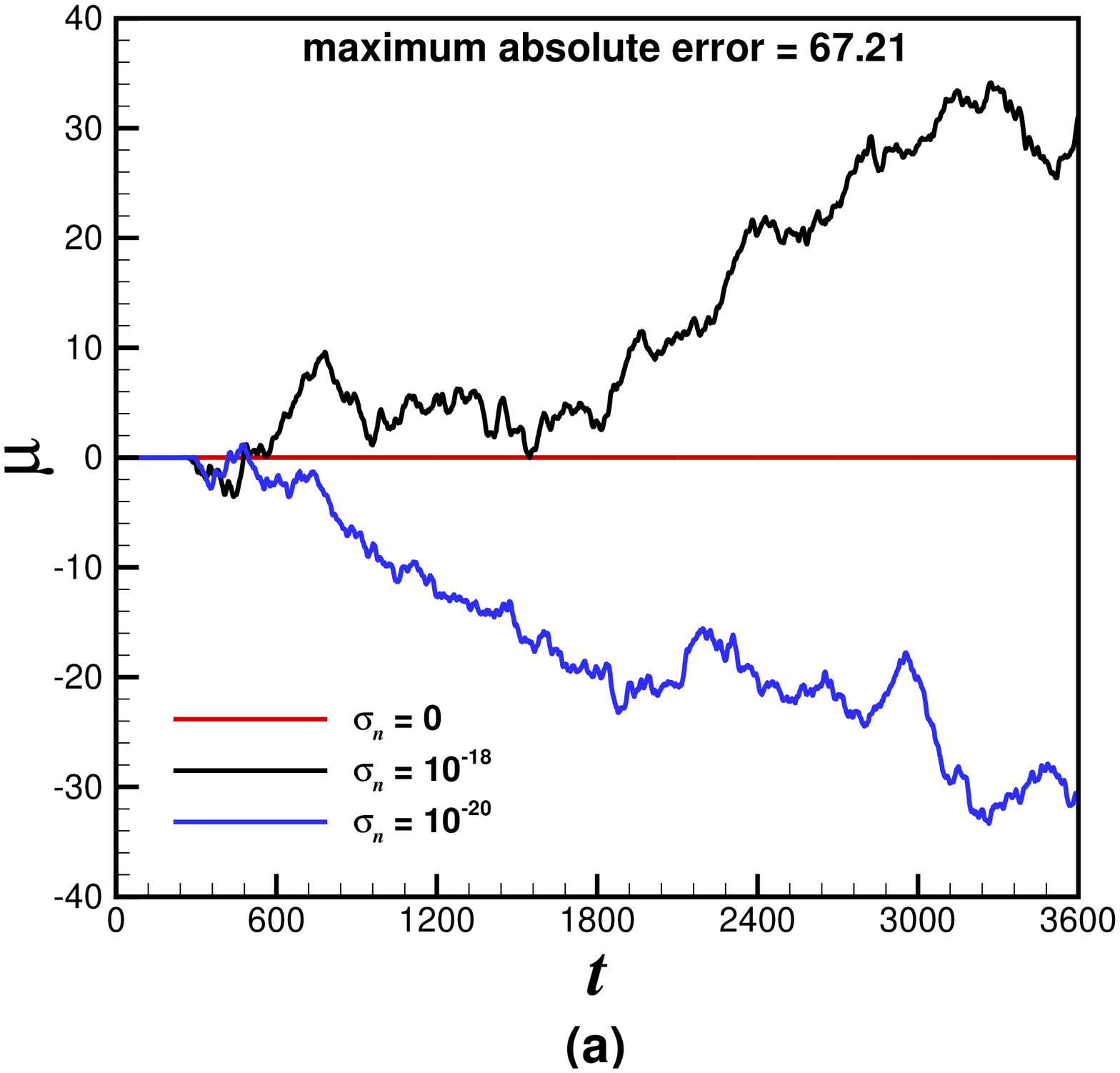}
  \includegraphics[width=2.5in]{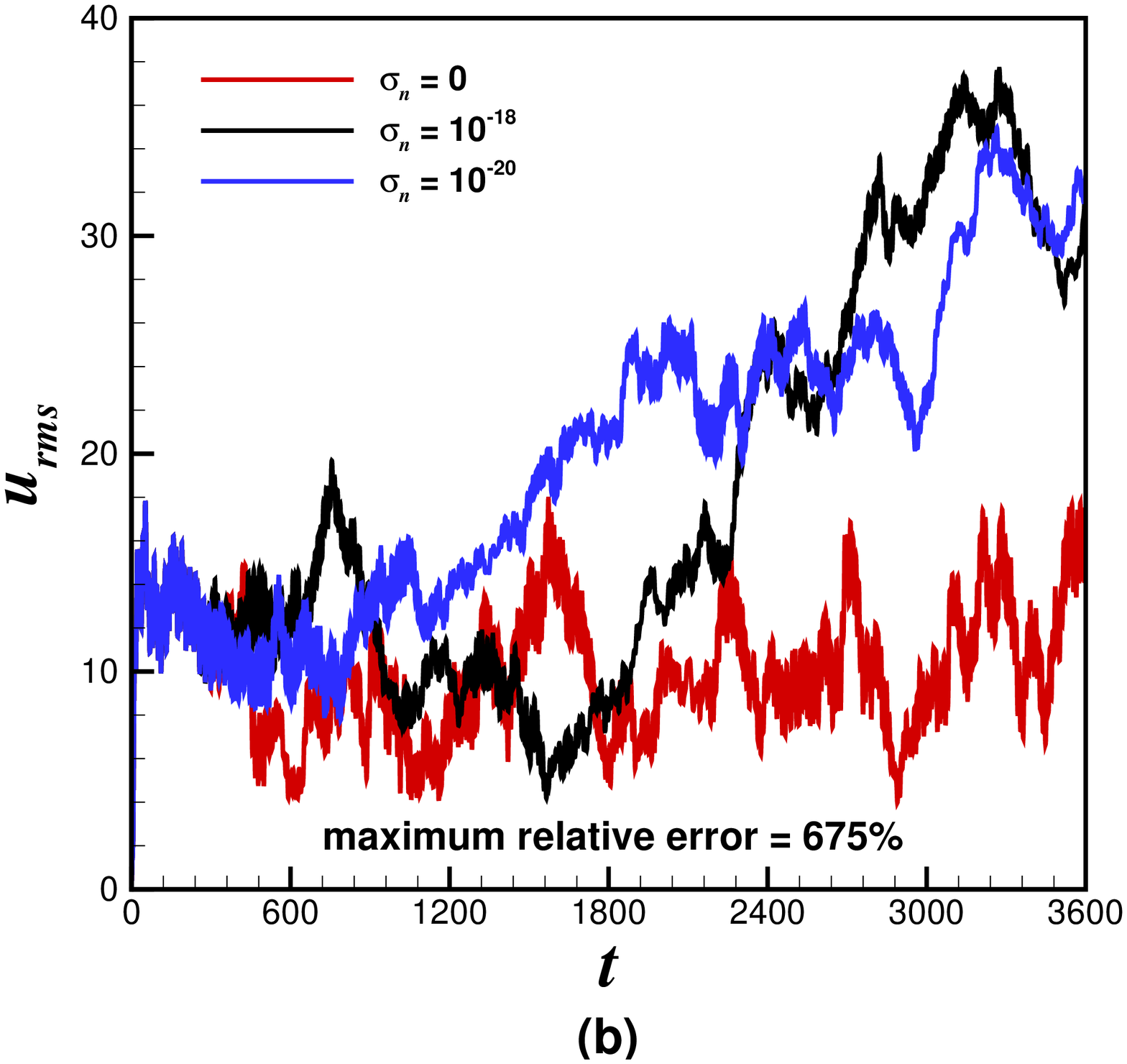}
  \caption{{\bf Influence of small disturbances to the spatial statistics of an ultra-chaos.}  (a) The spatial mean $\mu(t)$;  (b) The spatial Root-Mean-Square (RMS) $u_{rms}(t)$. The  curves are  based on the CNS results in $t\in[0,3600]$ of an ultra-chaos governed by the damped driven sine-Gordon equations (\ref{SGE})-(\ref{ini}) in case of  $\sigma_n=0$ (red), $\sigma_n=10^{-18}$ (black) and $\sigma_n=10^{-20}$ (blue), with the maximum absolute error $67.21$ for $\mu(t)$ and the maximum relative error $675\%$ for $u_{rms}(t)$, respectively, where $\sigma_{n}$ denotes the standard deviation of Gaussian white noise.}
  \label{fig-4}
\end{figure}

How about their statistic properties? Figure~\ref{fig-4} shows the influence of small disturbances to the spatial statistics, i.e. the spatial mean $\mu(t)$ and the  spatial root-mean-square (RMS) $u_{rms}(t)$, in case of $\sigma_n=0$, $\sigma_n=10^{-18}$ and $\sigma_n=10^{-20}$, respectively, with the maximum absolute error $67.21$ for $\mu(t)$ and the maximum relative error $675\%$ for $u_{rms}(t)$.  Note that, when there exist no environmental noises ($\sigma_n = 0$),  the spatial mean $\mu(t)$ is always zero and $u_{rms}(t)$ is always less than 20.  However,  the tiny environmental noise leads to a gradually increasing deviation of  $\mu(t)$  from zero and  results in $u_{rms}(t) > 20$ in $t\in[2200,3600]$.
Figure~\ref{fig-5} shows the influence of the small disturbances to  the temporal statistics, i.e. the temporal mean $\mu(x)$ and the temporal root-mean-square (RMS) $u_{rms}(x)$,  in case of $\sigma_n=0$, $\sigma_n=10^{-18}$ and $\sigma_n=10^{-20}$, respectively, with the maximum absolute error $33.98$ for $\mu(x)$ and the maximum relative error $187\%$ for $u_{rms}(x)$.  Note that, when there exist no environmental noises ($\sigma_n = 0$), we always have $|\mu(x)|<10$ and $u_{rms}(x)< 15$.  However, due to the tiny environmental noise, it mostly holds  $|\mu(x)| > 10$ and $u_{rms}(x) > 15$.
Note that, when there exist no environmental noises ($\sigma_n = 0$),  the {\em global} mean $\mu$ is always zero, but the tiny environmental noise leads to the huge deviations of the global mean $\mu=14.12$ in the case of  $\sigma_{n} = 10^{-18}$ and $\mu=-17.64$ in the case of  $\sigma_{n} = 10^{-20}$, respectively.

\begin{figure*}[t]
  \centering
  \includegraphics[width=2.5in]{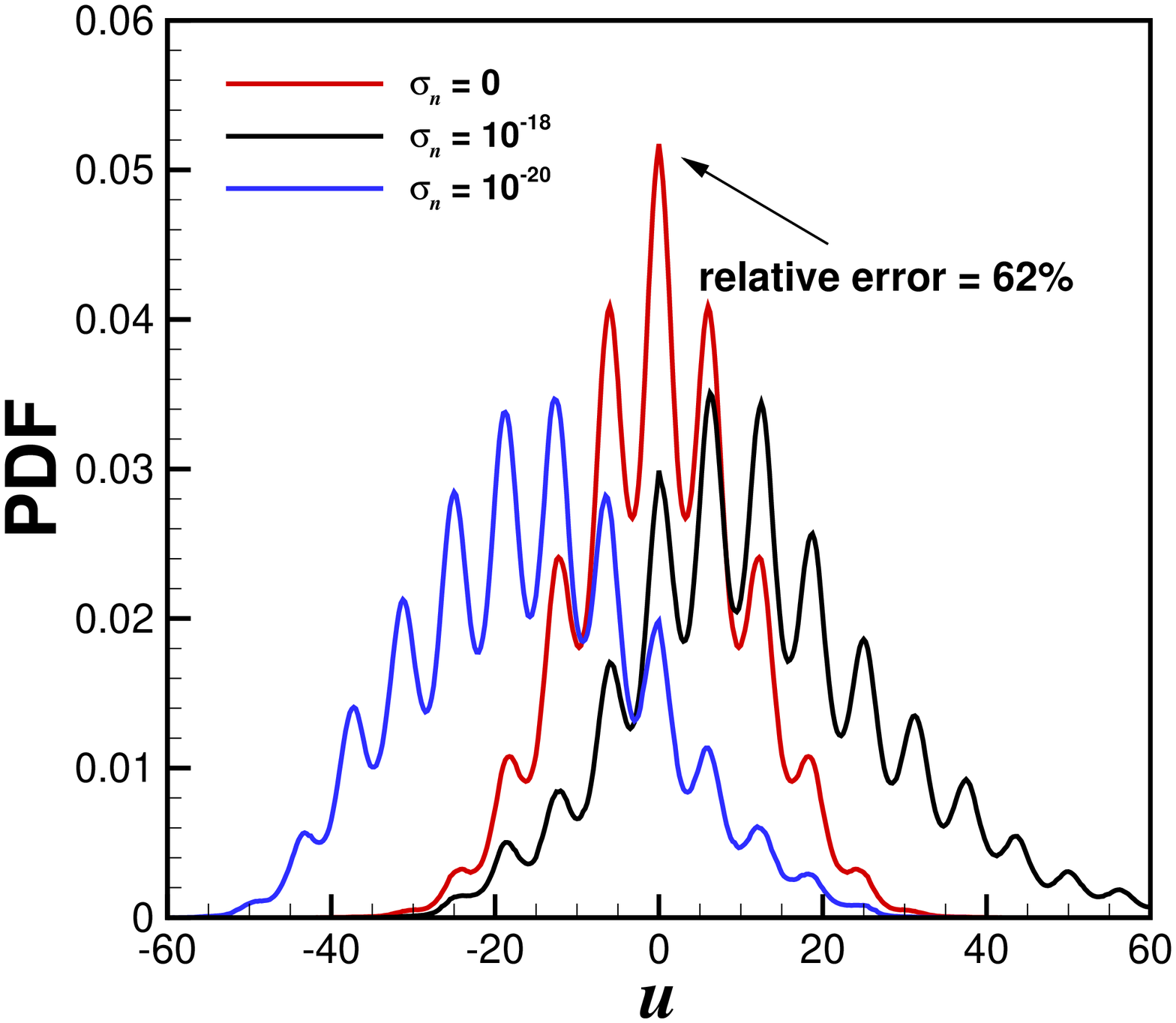}
  \caption{{\bf Influence of small disturbances to the probability density functions (PDFs)  of an ultra-chaos.}  The PDFs of $u(x,t)$  are  based on the CNS results in $t\in[0,3600]$ of an ultra-chaos governed by the damped driven sine-Gordon equations (\ref{SGE})-(\ref{ini}) in case of  $\sigma_n=0$ (red), $\sigma_n=10^{-18}$ (black) and $\sigma_n=10^{-20}$ (blue), respectively, where the relative error between the PDFs of $u = 0$ given by $\sigma_n=0$ and $\sigma_n=10^{-20}$ reaches $62\%$, and $\sigma_{n}$ denotes the standard deviation of Gaussian white noise.}
  \label{fig-6}
\end{figure*}

\begin{figure*}[!ht]
  \centering
  \includegraphics[width=2.5in]{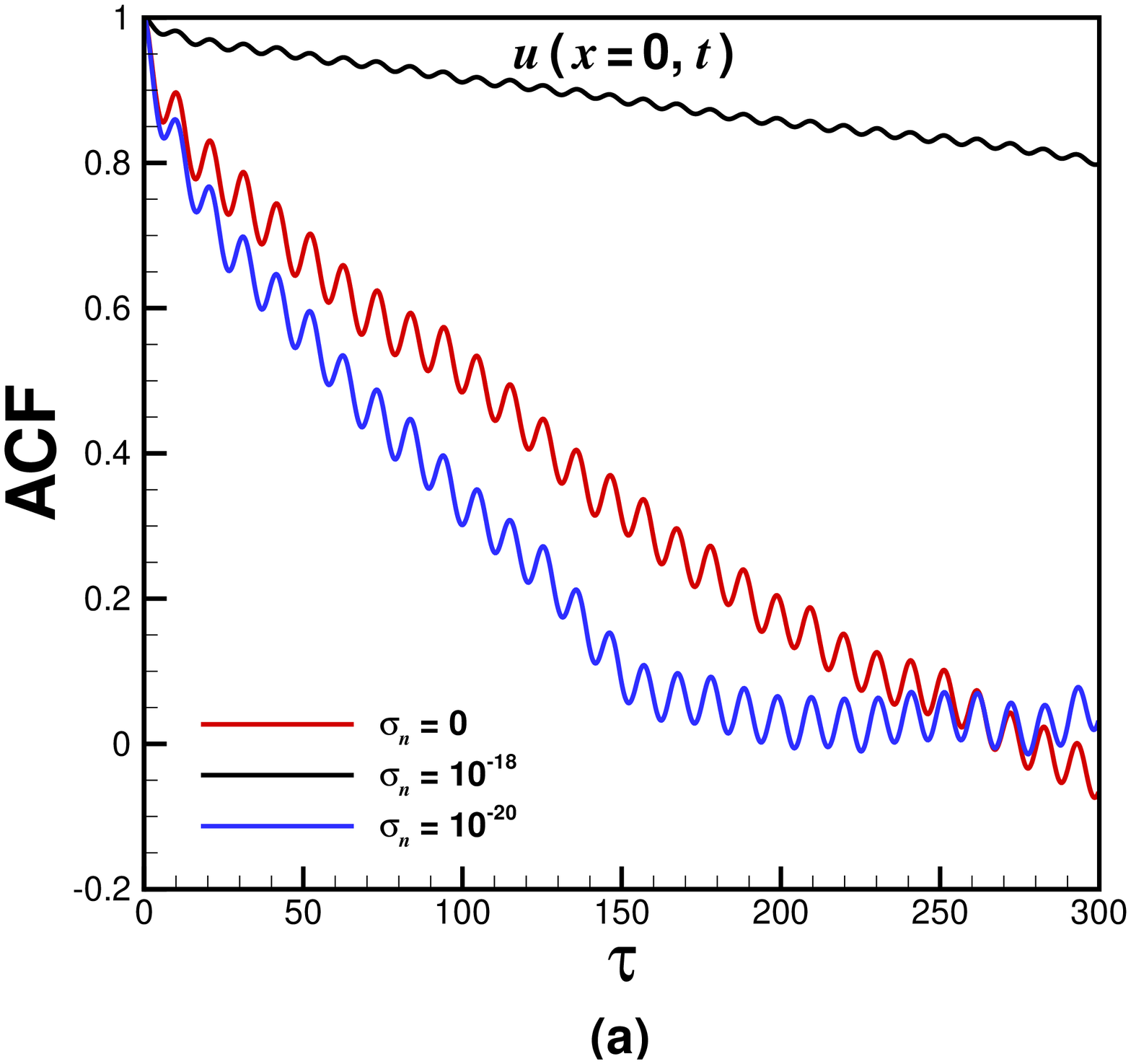}
  \includegraphics[width=2.5in]{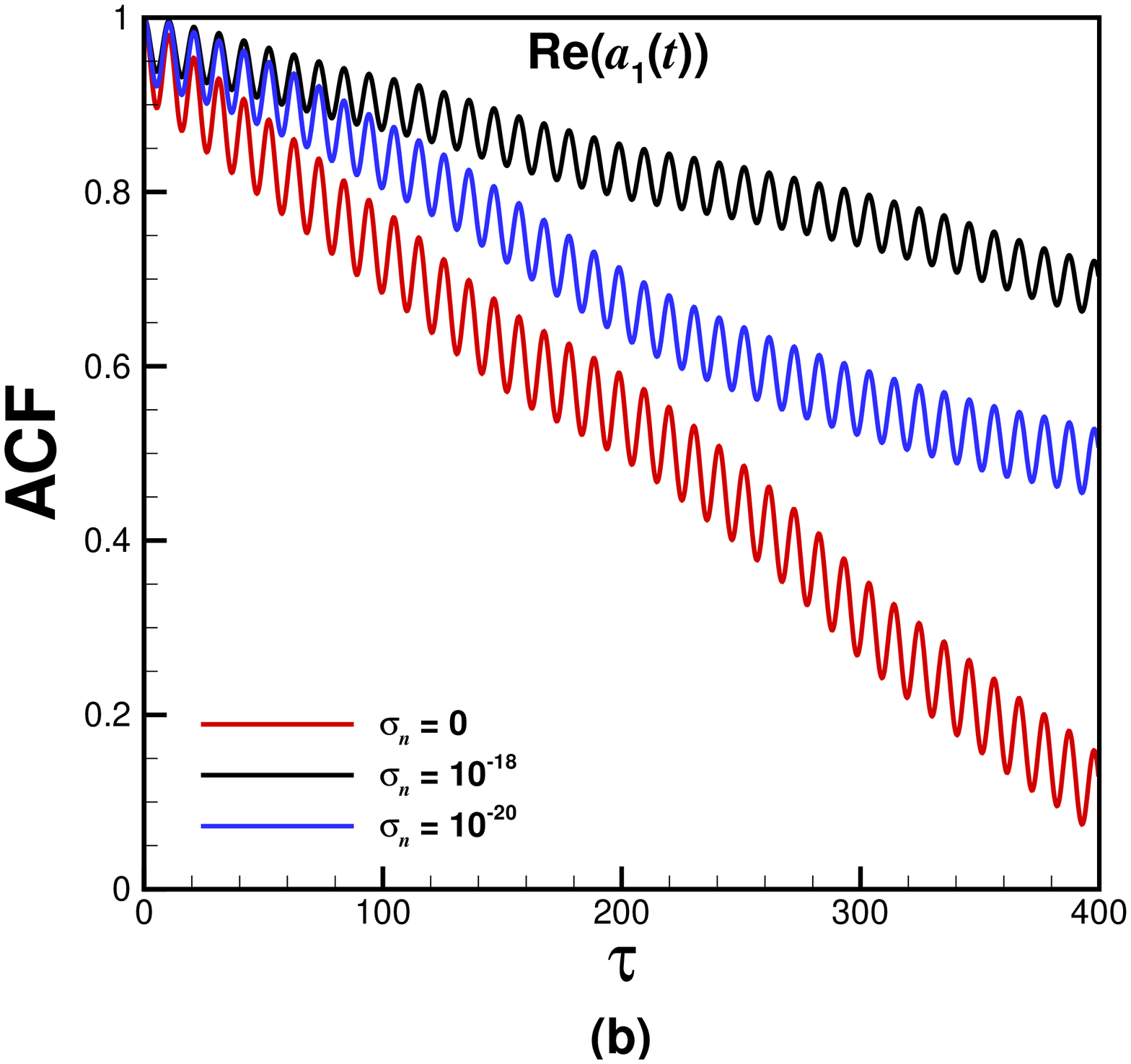}
  \caption{{\bf Influence of small disturbances to the autocorrelation functions (ACFs)  of an ultra-chaos.}  (a) ACF of $u(0,t)$;   (b) ACF of $Re(a_1(t))$.  The  curves are  based on the CNS  results $u(x,t)$ in $t\in[0,3600]$ of an ultra-chaos governed by the damped driven sine-Gordon equations (\ref{SGE})-(\ref{ini}) in case of $\sigma_n=0$ (red), $\sigma_n=10^{-18}$ (black) and $\sigma_n=10^{-20}$ (blue), respectively, where $Re(a_1)$ is the real part of the coefficient $a_{1}(t)$ of Fourier series (\ref{Fourier-series}) and $\sigma_{n}$ denotes  the standard deviation of Gaussian white noise.}
 \label{fig-7}
\end{figure*}

Figure~\ref{fig-6} shows the influence of small disturbances to the probability density functions (PDFs) of the chaotic simulation $u(x,t)$ in case of $\sigma_n=0$, $\sigma_n=10^{-18}$ and $\sigma_n=10^{-20}$, respectively, where the relative error of the PDFs at $u = 0$ in case of  $\sigma_n=0$ and $\sigma_n=10^{-20}$ reaches $62\%$.  Note that, when there exist no environmental noises ($\sigma_n = 0$),  the PDF has a kind of symmetry about $u=0$.  However, such kind of symmetry is lost due to appearance of the environmental noises.  In other words,  the environmental noises result in the {\em symmetry breaking} of the  PDFs of $u(x,t)$.    Besides,  the tiny difference between the external environmental noises (such as $\sigma_n = 10^{-18}$ and $\sigma_n = 10^{-20}$)  leads to huge deviation of their PDFs of $u(x,t)$, as shown in Figure~\ref{fig-6}.

Figure~\ref{fig-7}(a) shows the influence of small external disturbances to the autocorrelation functions (ACFs) of the time series $u(0,t)$ in case of $\sigma_n=0$, $\sigma_n=10^{-18}$ and $\sigma_n=10^{-20}$.  Obviously,   the ACF of  $u(0,t)$ has sensitivity dependance on the environmental noises (SDEN).  Especially,  the ACF of  $u(0,t)$  in case of  $\sigma_n = 10^{-18}$ has huge deviations from those in case of $\sigma_n= 10^{-20}$ and $\sigma_n= 0$.
 Besides, expand $u(x,t)$ in Fourier series
\begin{equation}
 u(x,t) =  \sum_{k=0}^{+\infty} a_k(t) \exp( k \lambda x {\bf i} ), \label{Fourier-series}
\end{equation}
where $a_k(t)$  denotes the amplitude and ${\bf i} = \sqrt{-1}$.    It is found that
the autocorrelation functions  (ACFs) of the amplitude $a_k(t)$ of the Fourier series of $u(x,t)$ also have sensitivity dependence on the environmental noises, too, as shown in Figure~\ref{fig-7}(b) for the  comparison of the ACFs of the real part of  $a_1(t)$ in case of $\sigma_n=0$, $\sigma_n=10^{-18}$ and $\sigma_n=10^{-20}$.

\begin{figure*}[t]
  \centering
  \includegraphics[width=2.7in]{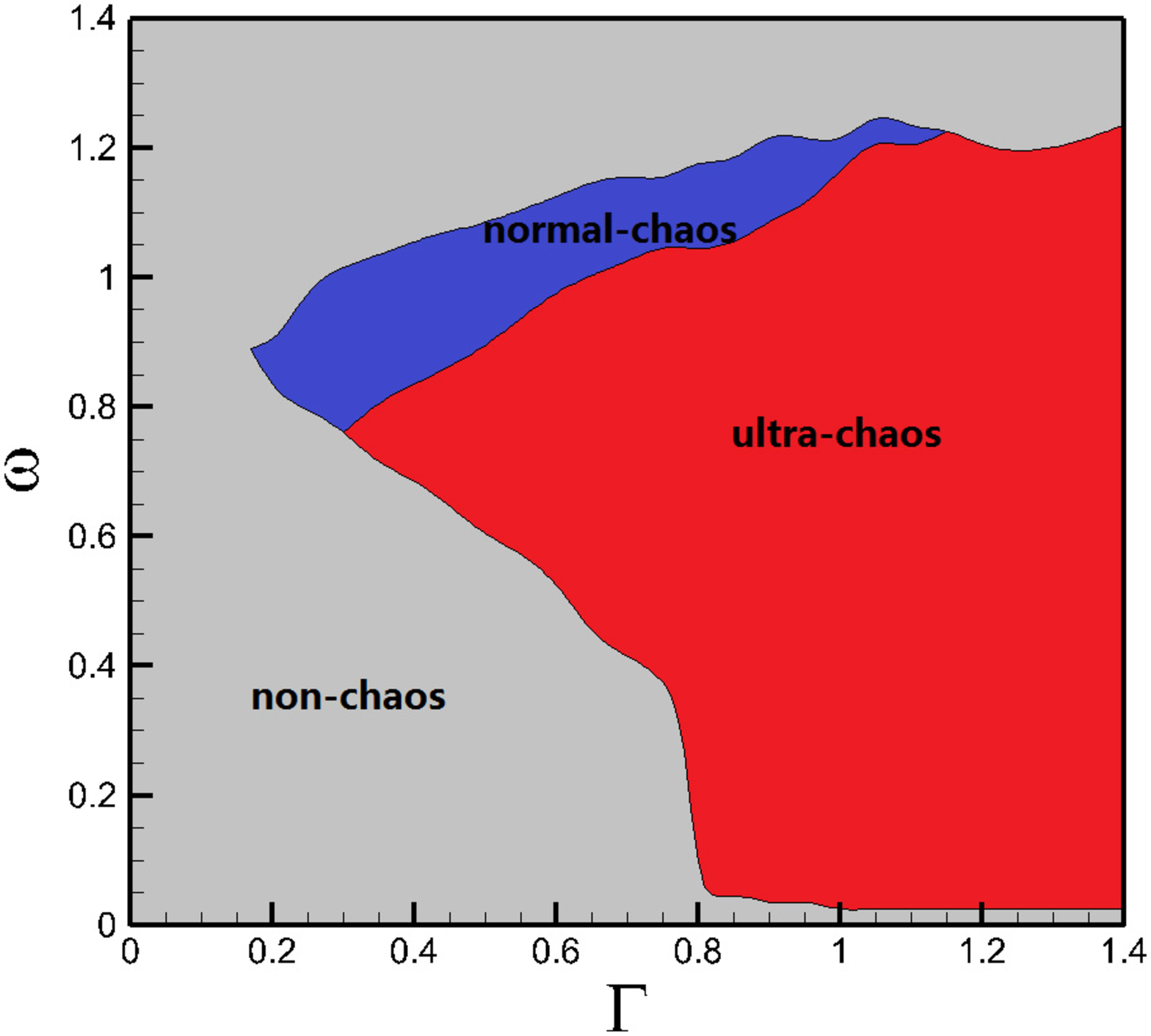}
  \caption{{\bf Phase plot of chaos of the damped driven  sine-Gordon equation (\ref{SGE}), (\ref{periodic}) and (\ref{ini}).} Red domain: ultra-chaos; Blue domain: normal-chaos; Gray domain: non-chaos (the state without any chaos). This phase plot is given by different values of the amplitude $\Gamma$ and temporal frequency $\omega$ of external force, with the constant $\alpha=1/10$ related to the damped friction, the total length $l=500$ of the system and the spatial frequency $\lambda = 2\pi/l$.}
  \label{fig-8}
\end{figure*}

\begin{figure*}[t!]
    \begin{center}
             \includegraphics[width=2.5in]{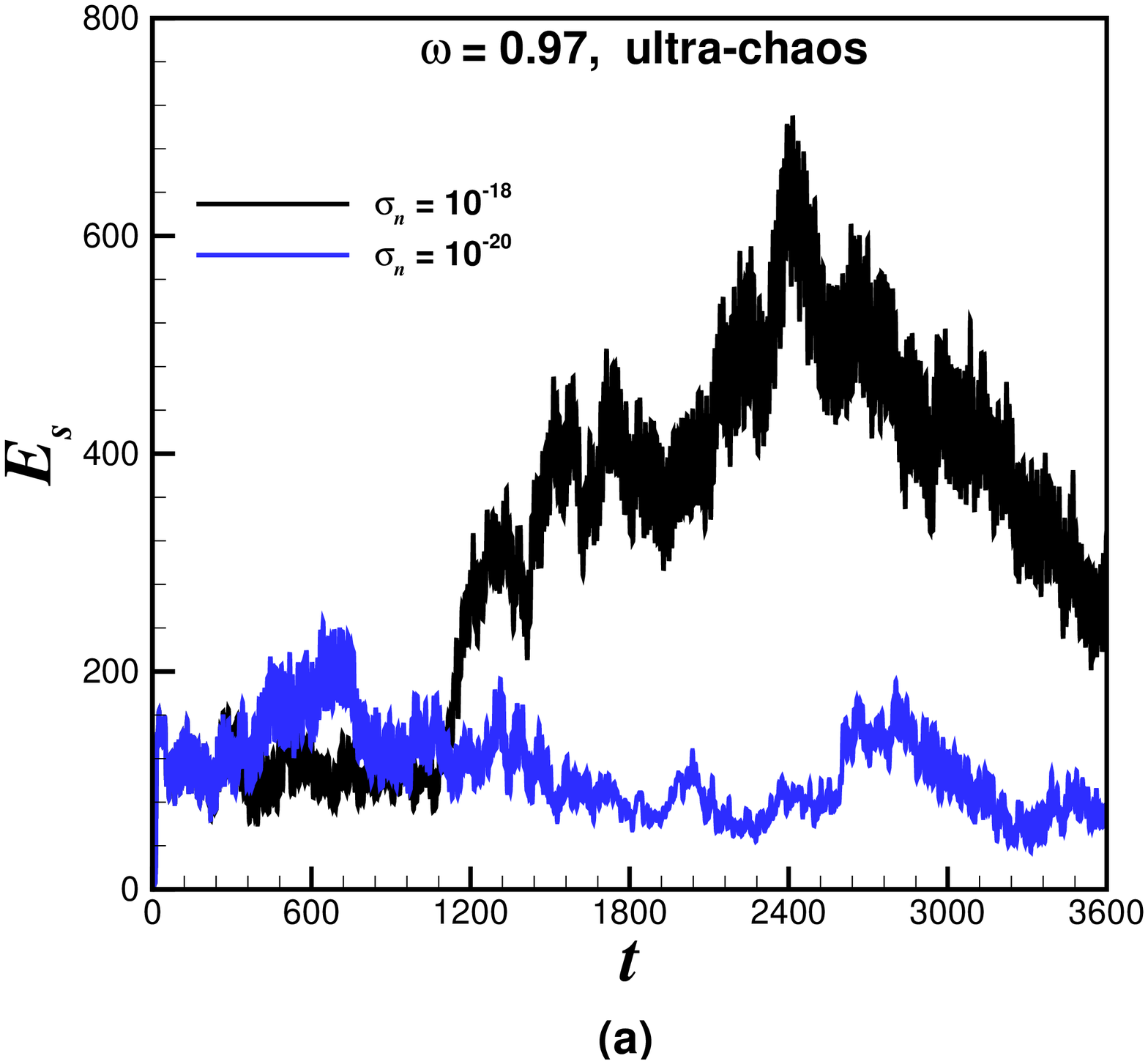}
             \includegraphics[width=2.5in]{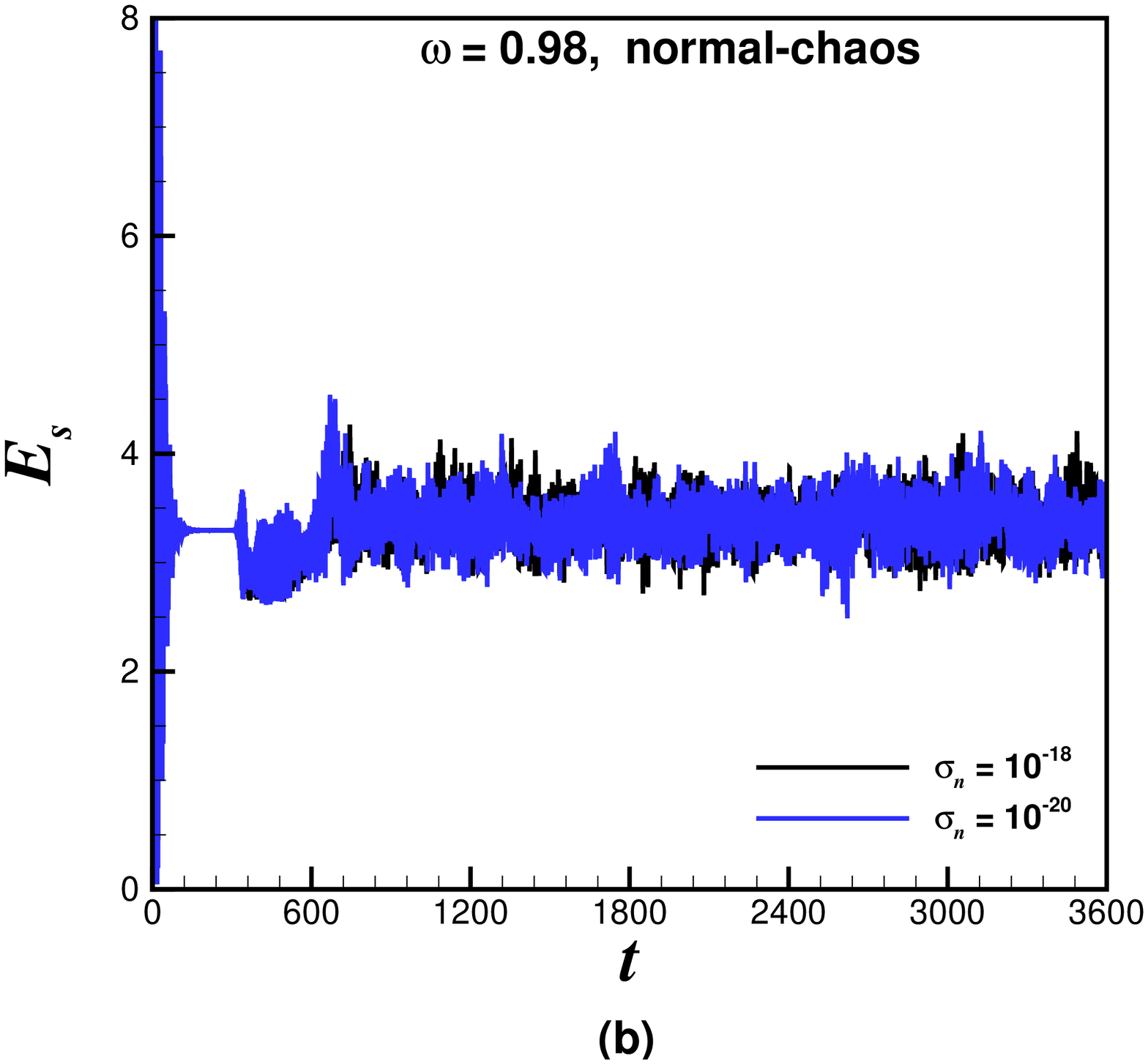}\\
             \includegraphics[width=2.5in]{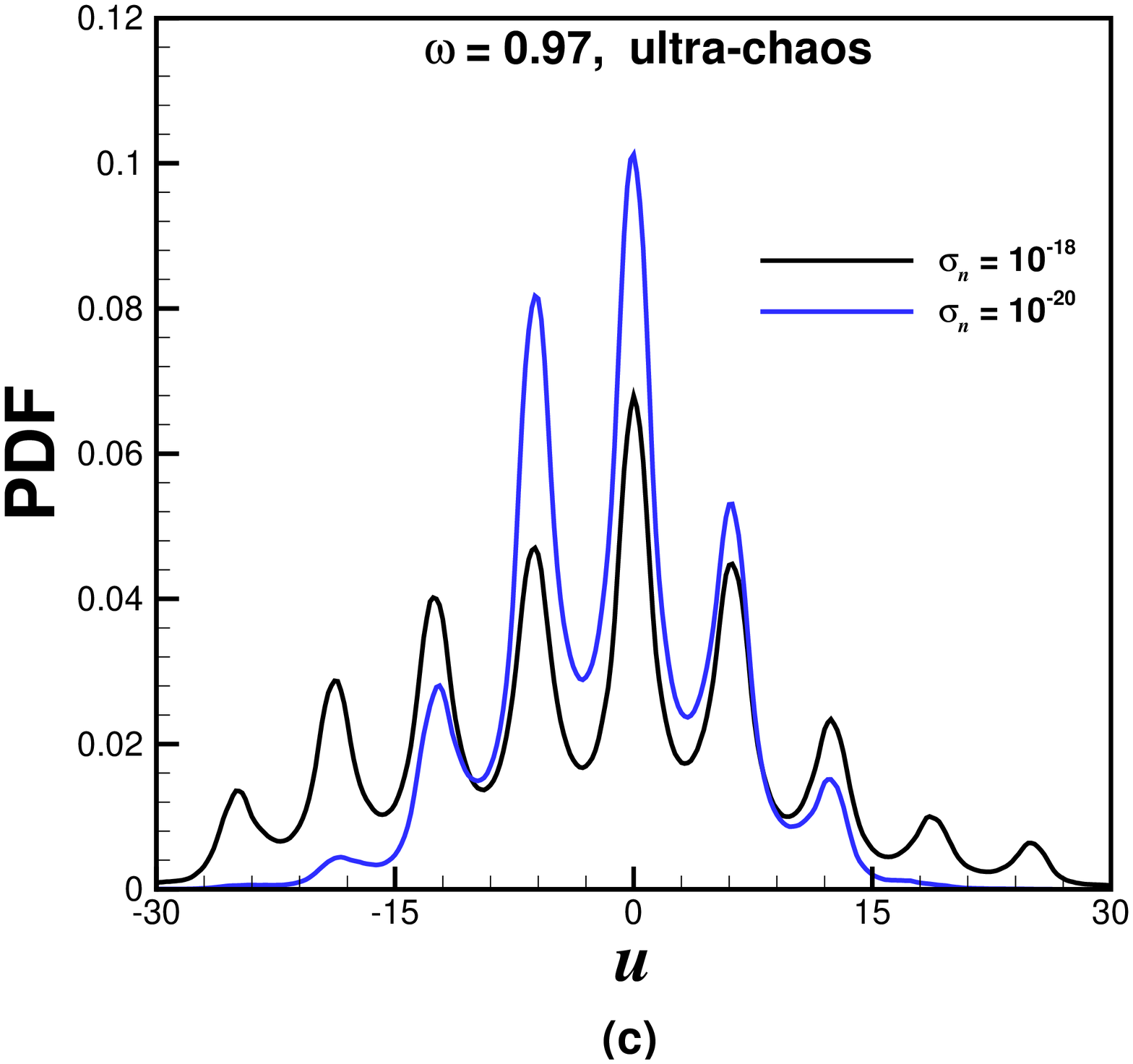}
             \includegraphics[width=2.5in]{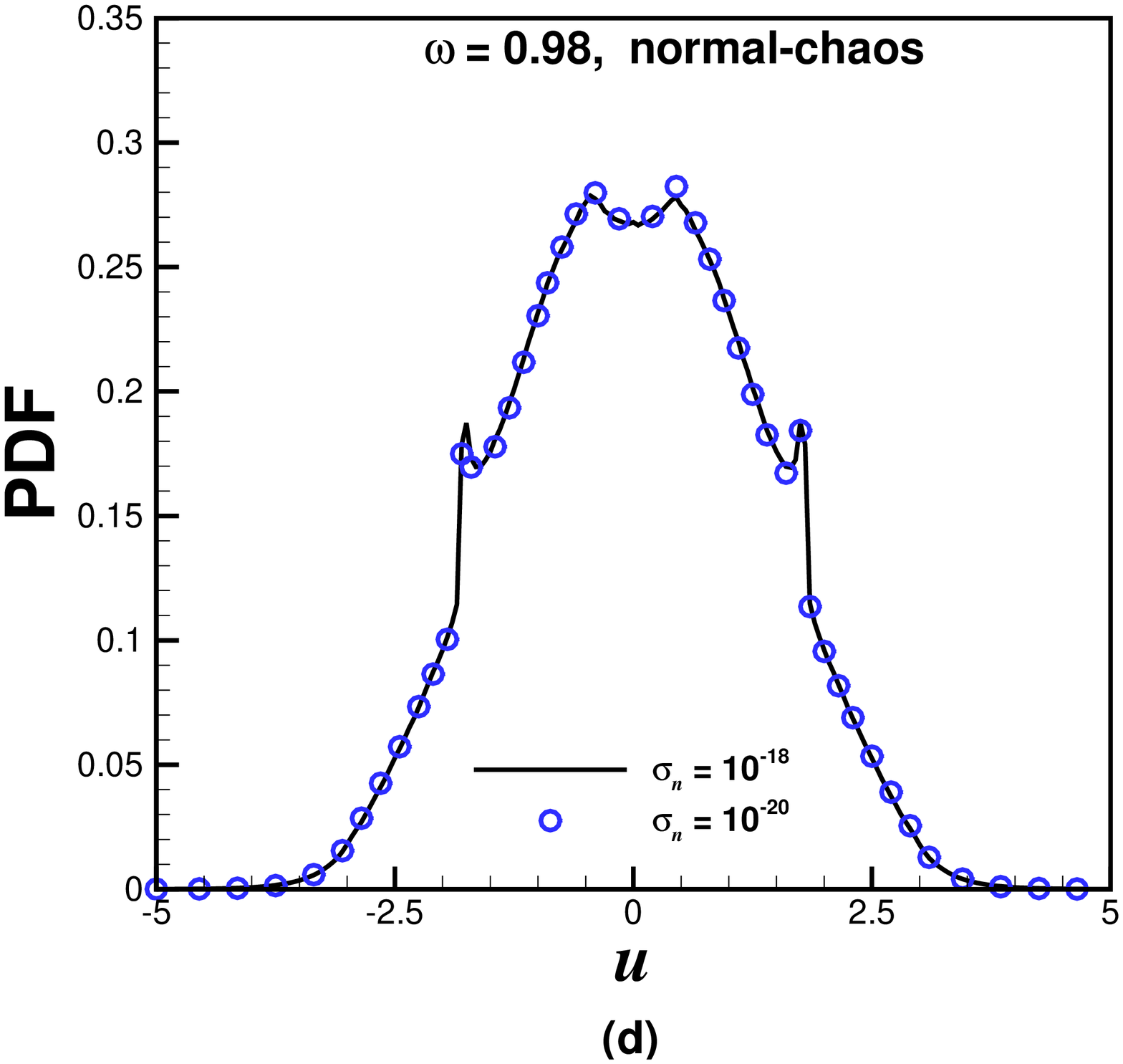}
    \caption{ {\bf Influence of tiny external noises to the statistics of an ultra-chaos  and a normal-chaos.} (a) and (b): the total spectrum energy $E_s(t)$; (c) and (d): the probability density function (PDF). The curves are based on the CNS results in $t\in[0,3600]$ of an ultra-chaos (a) and (c) in case of $\omega=0.97$ (with the maximum Lyapunov exponent $\lambda_{max}=0.13$) and a normal-chaos (b) and (d) in case of $\omega=0.98$ (with the maximum Lyapunov exponent $\lambda_{max}=0.016$) governed by the damped driven sine-Gordon equations (\ref{SGE}), (\ref{periodic}) and (\ref{ini}) under the two tiny external disturbances with the standard deviation $\sigma_n=10^{-18}$ (black) and $\sigma_n=10^{-20}$ (blue) of Gaussian white noises, respectively, when $\Gamma=0.6$ and the other parameters are the same as (\ref{parameters}). }
     \label{fig-9}
    \end{center}
\end{figure*}

Therefore, the tiny external  disturbances indeed lead to the great deviations of chaotic simulations of the damped driven sine-Gordon equation {\em not only} in trajectories {\em but also} in statistics.   In other words, statistical results of an ultra-chaos are {\em sensitive} to small external disturbances.

Does such kind of ultra-chaos {\em widely} exist?  Note that the above-mentioned ultra-chaos is found by using the physical parameters (\ref{parameters}).  Using various  amplitude $\Gamma$ and temporal frequency $\omega$ of external force, while keeping other physical parameters unchanged,  we find that the so-called ultra-chaos widely exists for the damped driven  sine-Gordon equation (\ref{SGE}), as shown in Figure~\ref{fig-8}.   Note that,  when $\Gamma = 0.6$ and $\omega$ increases from 0 to 1.4,   the system transfers from non-chaos to ultra-chaos, then from  ultra-chaos to normal-chaos, and final from normal-chaos to non-chaos.  
For example, let us consider the two cases of $\omega=0.97$ and $\omega=0.98$ when $\Gamma = 0.6$, where other physical parameters are the same as (\ref{parameters}).   Note that, when $\omega=0.98$, the total spectrum energy $E_{s(t)}$ and the PDFs are {\em not} sensitive to the tiny external disturbances, corresponding to a normal-chaos (with the maximum Lyapunov exponent $\lambda_{max}=0.016$), as shown in Figure~\ref{fig-9}.  However, when  $\omega = 0.97$, the total spectrum energy $E_{s(t)}$ and especially the PDFs are rather {\em sensitive} to the very tiny external disturbances, corresponding to an ultra-chaos  (with the maximum Lyapunov exponent $\lambda_{max}=0.13$).    Thus, as $\omega$ decreases from 0.98 to 0.97,  the system suddenly changes from a normal-chaos to an ultra-chaos.  This example illustrates that an ultra-chaos is indeed  quite different from a normal-chaos in essence.          
All of these also illustrate that the so-called ultra-chaos indeed widely exist and thus has general meanings.

This is a surprising result.  Such kind of chaos with statistical sensitivity on tiny external disturbances has never been reported, to the best of our knowledge.    Note that the four-dimensional  R\"{o}ssler equations \cite{Stankevich2020Chaos}  with two positive Lyapunov exponents belongs to the so-called hyper-chaos \cite{Rossler1979,Eiswirth1992CPL,Kapitaniak1995CSF,Baier1995PRE,Stankevich2020Chaos},  i.e. a chaos with at least two positive Lyapunov exponents. But, this kind of hyper-chaos is {\em not} an ultra-chaos: it belongs to normal-chaos.  This reveals that the ultra-chaos is  essentially different from the previously reported types of chaos and thus is indeed a new concept.

It is worth noting that the  environmental noises  considered in this article are at a quite tiny level, i.e. $10^{-18}$ and $10^{-20}$, which is at least two and four orders of magnitude smaller than the round-off error of traditional algorithms in double precision.   So, if traditional algorithms in single/double precision are used,  the  tiny  environmental disturbances are  submerged by the man-made numerical noises  so that we can {\em not} investigate the  influence of these  tiny objective disturbances at all.   Note that the background numerical error (i.e. the round-off error and the truncation error) of the CNS algorithm used here for the damped driven  sine-Gordon equations (\ref{SGE})-(\ref{ini}) is at the level of $10^{-230}$,  which is at least 210 orders of magnitude smaller than the tiny environmental noises!   Note that any results given by the CNS are not {\em absolutely}  ``clean'',  since they also  contain  numerical noises.  However,  in a finite interval $t\in[0,T_{c}]$ of time,  the numerical noises of CNS results are much smaller  than all physical variables under considerations and thus are negligible, where $T_{c}$ is the so-called critical predictable time.  This is similar to the drinking water from waterworks: although it always contains few bacteria,  it is safe to the health of human-being as long as the number of bacteria (per unit volume)  is less than a standard.  Thus, all results given the  clean numerical simulation (CNS) are relatively ``clean''.   This reveals the importance of the critical predictable time $T_{c}$ as a key concept in the frame of the CNS.  So, it is the CNS which provides us a powerful tool to accurately investigate the influence of such tiny environmental noises to an ultra-chaos.

\begin{figure*}[t!]
    \begin{center}
             \includegraphics[width=2.5in]{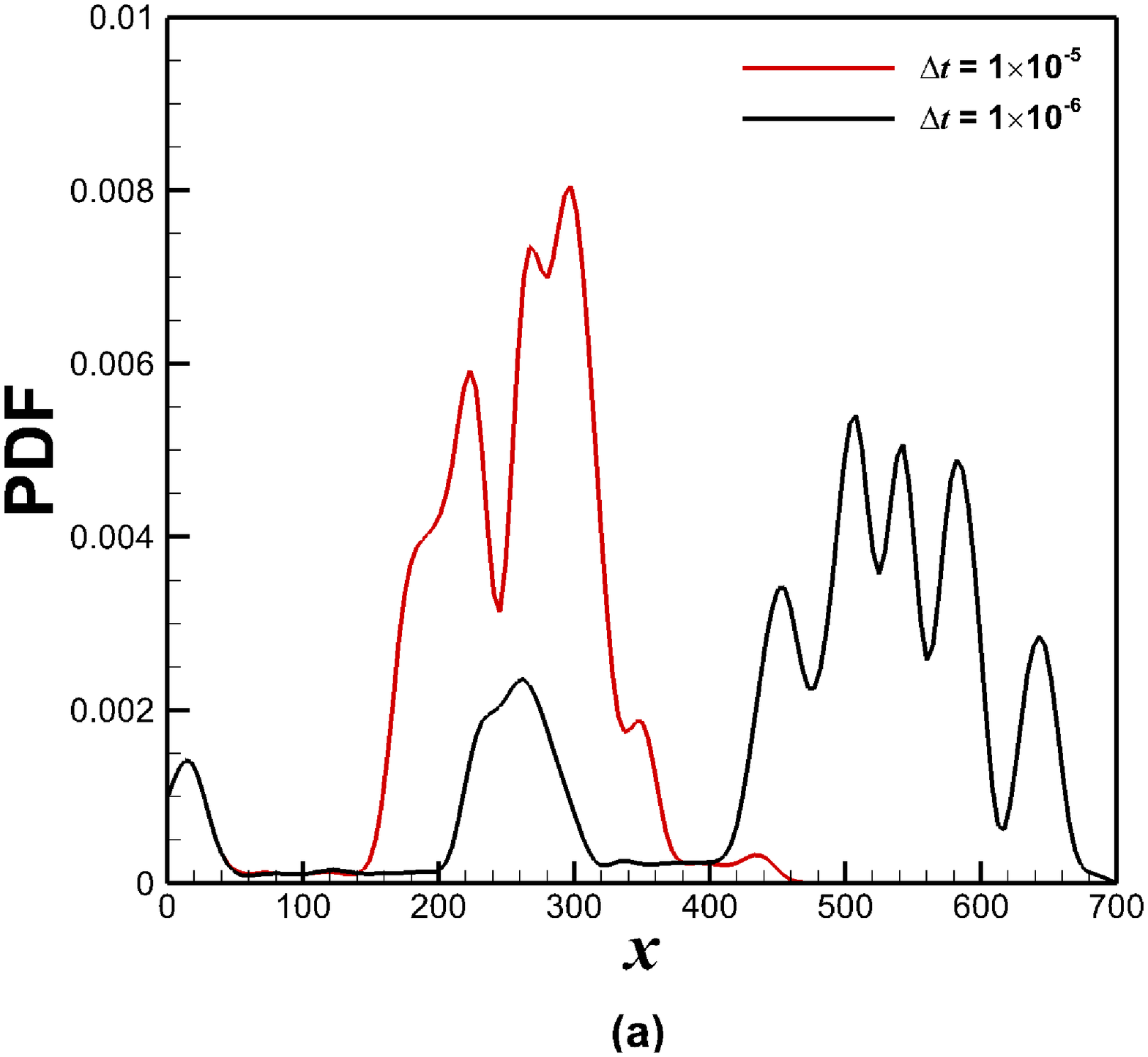}
             \includegraphics[width=2.5in]{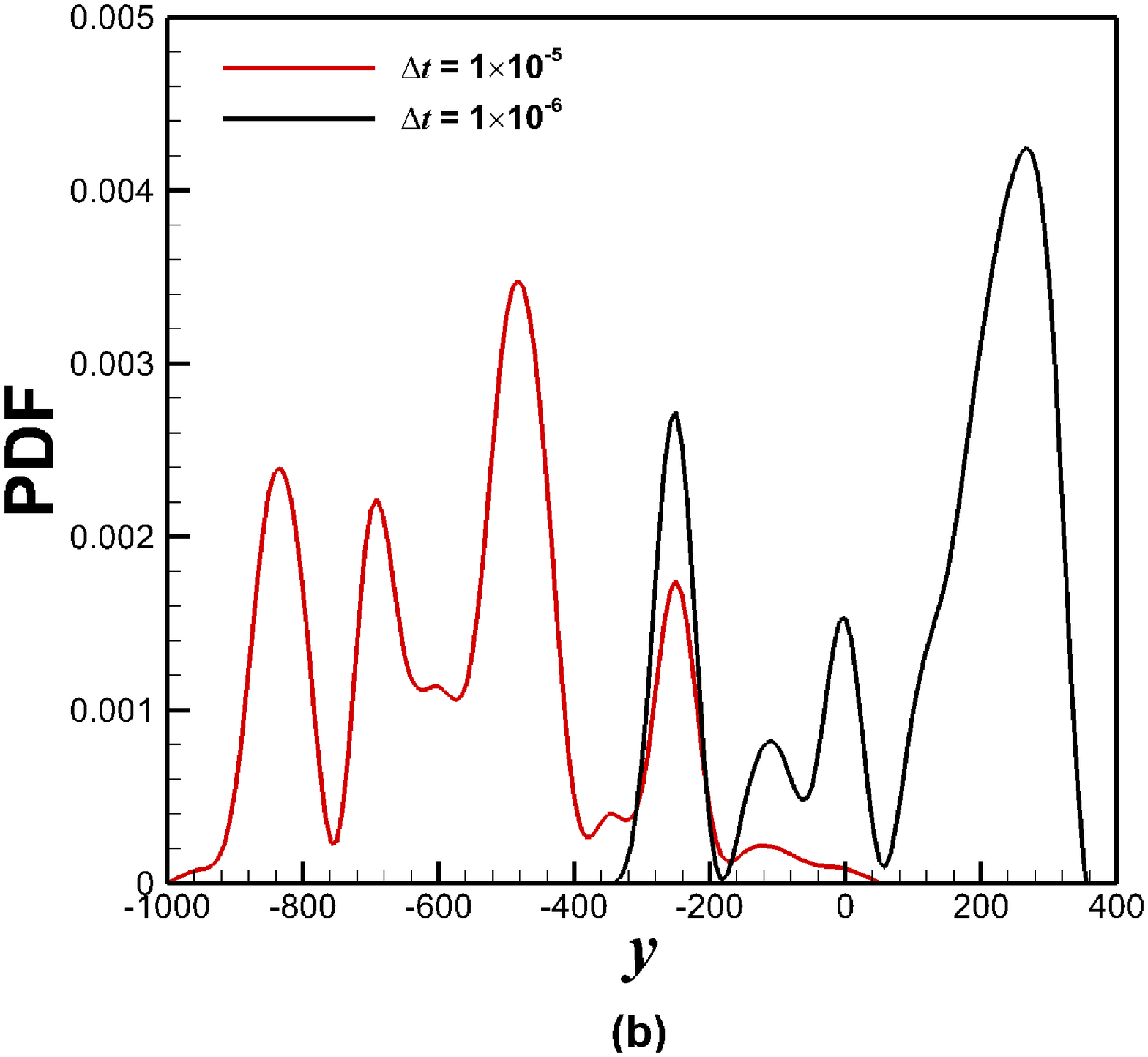}
    \caption{{\bf Influence of tiny numerical noises to the probability density functions (PDFs) of the ultra-chaos governed by (\ref{Arnold_x})-(\ref{Arnold_ini}) in case of $A=1$, $B=0.5$ and $C=0.5$.} (a) The PDFs of $x(t)$; (b) The PDFs of $y(t)$. The PDFs are based on the numerical results in $t\in[0, 10000]$ of Eqs.~(\ref{Arnold_x})-(\ref{Arnold_ini}),  given by the 4th-order Runge-Kutta's method with $40$-digit multiple-precision for all parameters and variables using two different time-steps $\Delta t = 10^{-5}$ and $\Delta t = 10^{-6}$, respectively.}
     \label{fig-10}
    \end{center}
\end{figure*}

To show the wide existence of the so-called ultra-extra, let us further consider the following ODEs: 
\begin{align}
&  \dot{x}=A\, \sin(z)+C\, \cos(y),    \label{Arnold_x}    \\
&  \dot{y}=B\, \sin(x)\, \tanh^{2}(x)+A\, \cos(z),    \label{Arnold_y}    \\
&  \dot{z}=C\, \sin(y)+B\, \cos(x)    \label{Arnold_z}
\end{align}
 with the initial condition
\begin{equation}
x(0)=y(0)=z(0)=0.       \label{Arnold_ini}
\end{equation}
 In case of $A=1$, $B=0.5$ and $C=0.5$,  the above equations have chaotic simulations with one positive Lyapunov exponent $\lambda=0.04$.   We gain the numerical simulations of (\ref{Arnold_x})-(\ref{Arnold_ini}) in a quite long interval $t\in[0,10000]$ by means of the 4th-order Runge-Kutta's method with $40$-digit multiple-precision for all parameters and variables but using two different time-steps $\Delta t = 10^{-5}$ and $\Delta t = 10^{-6}$, respectively.   Figure~\ref{fig-10} shows the influence of tiny numerical noises to the probability density functions (PDFs) of the chaotic simulation $x(t)$ and $y(t)$ in case of  $\Delta t = 10^{-5}$ and $\Delta t = 10^{-6}$ (whose numerical noises are at the level of $10^{-20}$ and $10^{-24}$), respectively.   Obviously, the PDFs of $x(t)$ and $y(t)$ are rather sensitive to the tiny numerical noises.    Note that  these numerical noises are equivalent to tiny external noises.   Thus,  Eqs.~(\ref{Arnold_x})-(\ref{Arnold_ini})  in case of   $A=1$, $B=0.5$ and $C=0.5$ correspond to an ultra-chaos.  This illustrates once again that ultra-chaos should exist widely.     

In summary,  the above-mentioned  examples illustrate that small disturbances can indeed lead to the huge deviations of an ultra-chaos,  not only in trajectory but also in {\em statistics}.   In other words,   statistical properties of an ultra-chaotic system have sensitivity dependance on small disturbances.  Therefore, even in {\em statistical} meanings,  small disturbances could result in the loss of the  reproducibility and replicability of an ultra-chaos,   no matter what and how we do.   Obviously, an ultra-chaos is at the higher-level of disorder than a normal-chaos.   It should be emphasized that small external disturbances always exist in practice, and are out of control.  Thus,  theoretically speaking,  the reproducibility and replicability  for an ultra-chaos are essentially impossible in practice, {\em forever}.  Therefore, ultra-chaos is indeed an {\em insurmountable} objective obstacle of reproducibility and replicability.

\section{Concluding remarks and discussions}

In this paper, a new concept, i.e. ultra-chaos, is proposed for the first time.  Unlike a normal-chaos,  statistical properties such as the probability density functions (PDF) of an ultra-chaos are sensitive to tiny disturbances.  We illustrate that ultra-chaos is widely existed and thus has general scientific meanings.  Obviously,  for an ultra-chaotic system whose statistics are sensitive to small disturbances,  it  is impossible in practice to replicate  any discoveries  based  on  ``statistically significant'' findings, even in the statistical meanings.  Obviously,  an ultra-chaos is indeed an  {\em insurmountable}  obstacle to  reproducibility and replicability.  The ultra-chaos as a new concept might open a new door and possibility to study chaos theory, turbulence theory,  computational fluid dynamics (CFD),  the statistical significance, reproducibility crisis,  and so on. 

First, given a chaotic system, we must judge whether it is an ultra-chaos or not.  For a normal-chaos, it is even {\em unnecessary} to use a very fine grid to gain its numerical simulations, since its statistics are {\em not} sensitive to numerical noises.     Besides, it is also {\em unnecessary} to consider the influence of  small environmental noises to a normal-chaos, since environmental noises are essentially  equivalent to numerical noises in mathematics.    However, for an ultra-chaos,  its statistics are sensitive to tiny environmental noises, as illustrated here.  Note that every dynamical system has environmental noises, which are  random, unavoidable, and out of control.    Therefore, from practical viewpoint,  it is {\em impossible} to replicate an ultra-chaotic system even in the statistical meanings.  Theoretically speaking, it  is  impossible to make  any  reliable/replicable  predictions  about  such kind of ultra-chaos, {\em forever}, even in statistical meanings.  

Secondly, since turbulence has a close relationship with spatiotemporal chaos \cite{Henriksen1994},   it is important to know whether  a turbulent flow belongs to ultra-chaos or not.   If a turbulent flow is a kind of normal-chaos, which we designate as ``normal-turbulence'',  it is {\em unnecessary} to use a very fine grid of  discretization  for numerical simulations, since its statistics are not sensitive to numerical noises.   In this case,  numerical noise itself is not a trouble at all.    However, If a turbulent flow is a kind of ultra-chaos, which we designate as ``ultra-turbulence'',  it should be {\em impossible} in practice to gain  reproducible/replicable  results even in statistics, no matter how fine  grid  of  discretization for numerical simulation or how advanced  apparatus with very high precision for experimental measures are used.   This also reveals the importance of ultra-chaos as a new concept, since turbulence is a very important field in science and engineering.

Thirdly, the ultra-chaos as a new concept  should be of benefit to deepen our understanding about the so-called ``crisis of reproducibility''.
According to Nature's survey of 1,576 researchers in physics \& engineering, chemistry,  earth \& environment, biology, medicine  and so on, more than half of these surveyed have failed to reproduce their own experiments and agree that there indeed exists a significant ``crisis of reproducibility''  \cite{Monya2016Nature}.   In addition, in a current investigation on evaluating replicability of laboratory experiments in economics, the replication ratio is only 66\% on average \cite{Colin2016Science}.  The so-called ``crisis of reproducibility''  widely rises in numerous fields of science and engineering.  Currently,  computational science has led to exciting advances in many fields, such as the Computational Fluid Dynamics (CFD).    The reproducibility and replicability in the CFD serve as a minimum standard to  judge scientific claims and discoveries \cite{Peng2011Science}.   However, as reported by Mesnard and Barba \cite{Mesnard2017OSE}, ``completing a full replication study of previously published findings on bluff-body aerodynamics is harder than it looks, even when using good reproducible-research practices and sharing code and data openly''.    In a survey about the reproducible researches \cite{Peng2011Science}, the replication ratio  in computational science is only about 25\%.  There might be lots of reasons that could lead to irreproducible researches, such as  poor experimental design,  low statistical power,  poor analysis, insufficient oversight/mentoring, selective reporting, raw data not available from original lab,  code \& computational mesh unavailable, insufficient peer review, fraud, and so on \cite{Monya2016Nature,Mesnard2017OSE}.
The non-reproducibility of many scientific researches  has leaded to growing worry  about the reliability of claims of new  discoveries that are based on ``statistically significant'' findings \cite{Benjamin2018NatureHB}.
  For an ultra-chaos, the derivations caused  by either the artificial  disturbances  or  objective environmental noises are macroscopic, and more importantly, the corresponding  statistics are {\em sensitive} to these disturbances/noises.   Note that the statistical significance and $p$-values are  based on reliable/replicable results of statistics.  Naturally,  for an ultra-chaotic system whose statistics are sensitive to small disturbances,  it  should be impossible in practice to replicate  new  discoveries  based  on  ``statistically significant'' findings.  Without doubt,  an ultra-chaos is indeed an  {\em insurmountable}  obstacle to  reproducibility and replicability, and is really an {\em objective} cause   to ``crisis of reproducibility''.   

 What we would like to emphasize here is that, even if  {\em all} man-made disturbances could be smoothed out,  the reproducibility and replicability might be still {\em essentially} impossible even in statistical meanings for some dynamic systems such as an ultra-chaos.  Thus, in theory,  the ultra-chaos as a new concept reveals a kind of incompleteness.  
Similar to G\"{o}del's incompleteness theorem, such kind of  ``incompleteness of reproducibility and replicability'' reveals a {\em limitation} of our scientific paradigm based on reproducible/repeatable experiments (at least in statistics),  which can be traced back to Galileo.   Note that  scientific findings are based on the reproducibility and replicability of their main supporting evidences at least in statistics:  such kind of the reproducibility and replicability  is a cornerstone of the modern science.   It is an open question whether or not this  ``incompleteness of reproducibility and replicability''   might  shake  the cornerstone of  our modern science. 

Unlike the so-called hyper-chaos \cite{Rossler1979, Eiswirth1992CPL, Kapitaniak1995CSF, Baier1995PRE, Stankevich2020Chaos}, i.e. a chaotic system with at least two positive Lyapunov exponents, the characteristic of an ultra-chaos is the sensitivity dependence of statistics on  small disturbance.   Note that the hyper-chaos governed by the four-dimensional  R\"{o}ssler equation \cite{Stankevich2020Chaos}  is just only a normal-chaos.   This reveals that ultra-chaos  is essentially different from the previously reported types of chaos and thus is indeed a totally new concept.  

Finally, we emphasize that scientific claims should  gain credence due to  the reproducibility and replicability of their main supporting evidences.  Unfortunately,  we can {\em not} gain such kind of credence for an ultra-chaos!  Would this   ``incompleteness of reproducibility''   shake  the cornerstone of the modern science?   Would the ultra-chaos lead to a crisis of confidence in scientific researches?   How should  we  understand and interpret  numerical/experimental results of such kind of ultra-chaotic systems?    Should we regard random environmental noises as an inherent part of solution of an ultra-chaos?    How should we define the ``truth'' of an ultra-chaos?    What kind of  ``truth''  could an ultra-chaos  tell  us?   Do there exist some ultra-turbulence in practice?    Certainly, there are still lots of works to do in future.

\section*{Acknowledgments}
This work is partly supported by the National Natural Science Foundation of China (No. 91752104).



\begin{thebibliography}{99}
\bibitem{Poincare1890}
J.~H. Poincar{\' e}, {Sur} le probl{\' e}me des trois corps et les {\'
  e}quations de la dynamique. {D}ivergence des s{\' e}ries de m. {Lindstedt},
  Acta Mathematica 13 (1890) 1--270.

\bibitem{Lorenz1963}
E.~N. Lorenz, Deterministic nonperiodic flow, Journal of the Atmospheric
  Sciences 20~(2) (1963) 130--141.

\bibitem{Li1975Period}
T.~Y. Li, J.~A. Yorke, Period three implies chaos, American Mathematical
  Monthly 82~(10) (1975) 985--992.

\bibitem{Rossler1979}
O.~E. R\"{o}ssler, An equation for hyperchaos, Physics Letters A 71~(2--3)
  (1979) 155 -- 157.

\bibitem{Eckmann1985RMP}
J.~P. Eckmann, Ergodic theory of chaos and strange attractors, Reviews of
  Modern Physics 57~(3) (1985) 617 -- 656.

\bibitem{Lorenz1993Book}
E.~N. Lorenz, The Essence of Chaos, University of Washington Press, 1993.

\bibitem{PeterSmith1998Explaining}
P.~Smith, Explaining Chaos, Cambridge University Press, 1998.

\bibitem{Sprott2003Chaos}
J.~C. Sprott, {Chaos and Time-Series Analysis}, Oxford University Press, 2003.

\bibitem{sprott2010}
J.~C. Sprott, {Elegant Chaos: Algebraically Simple Chaotic Flows}, World
  Scientific, Singapore, 2010.

\bibitem{Christophe2021Chaos}
C.~Letellier, L.~F. Olsen, S.~Mangiarotti, Chaos: from theory to applications
  for the 80th birthday of {O}tto {E}. {R}\"{o}ssler, Chaos 31 (2021) 060402.

\bibitem{Lorenz2006}
E.~N. Lorenz, Computational periodicity as observed in a simple system, Tellus
  Series A-dynamic Meteorology \& Oceanography 58~(5) (2006) 549--557.

\bibitem{Teixeira2007}
J.~Teixeira, C.~A. Reynolds, K.~Judd, Time step sensitivity of nonlinear
  atmospheric models: Numerical convergence, truncation error growth, and
  ensemble design, Journal of the Atmospheric Sciences 64~(1) (2007) 175.

\bibitem{Liao2009}
S.~Liao, On the reliability of computed chaotic solutions of nonlinear
  differential equations, Tellus Series A-dynamic Meteorology \& Oceanography
  61~(4) (2009) 550--564.

\bibitem{Liao2013A}
S.~Liao, On the numerical simulation of propagation of micro-level inherent
  uncertainty for chaotic dynamic systems, Chaos, Solitons \& Fractals 47
  (2013) 1--12.

\bibitem{Liao2013B}
S.~Liao, Physical limit of prediction for chaotic motion of three-body problem,
  Communications in Nonlinear Science and Numerical Simulation 19~(3) (2014)
  601--616.

\bibitem{LIAO2014On}
S.~Liao, P.~Wang, On the mathematically reliable long-term simulation of
  chaotic solutions of {L}orenz equation in the interval [0, 10000], Science
  China Physics, Mechanics \& Astronomy 57~(2) (2014) 330--335.

\bibitem{li2014stability}
X.~Li, S.~Liao, On the stability of the three classes of {N}ewtonian three-body
  planar periodic orbits, Science China Physics, Mechanics \& Astronomy 57~(11)
  (2014) 2121--2126.

\bibitem{Lin2017On}
Z.~Lin, L.~Wang, S.~Liao, On the origin of intrinsic randomness of
  {R}ayleigh-{B}{\'e}nard turbulence, Science China Physics, Mechanics \&
  Astronomy 60~(1) (2017) 014712.

\bibitem{li2017more}
X.~Li, S.~Liao, More than six hundred new families of {N}ewtonian periodic
  planar collisionless three-body orbits, Science China Physics, Mechanics \&
  Astronomy 60~(12) (2017) 129511.

\bibitem{li2018over}
X.~Li, Y.~Jing, S.~Liao, Over a thousand new periodic orbits of a planar
  three-body system with unequal masses, Publications of the Astronomical
  Society of Japan 70~(4) (2018) 64.

\bibitem{li2019collisionless}
X.~Li, S.~Liao, Collisionless periodic orbits in the free-fall three-body
  problem, New Astronomy 70 (2019) 22--26.

\bibitem{Hu2020JCP}
T.~Hu, S.~Liao, On the risks of using double precision in numerical simulations
  of spatiotemporal chaos, Journal of Computational Physics 418 (2020) 109629.

\bibitem{Qin2020CSF}
S.~Qin, S.~Liao, Influence of numerical noises on computer-generated simulation
  of spatiotemporal chaos, Chaos, Solitons \& Fractals 136 (2020) 109790.

\bibitem{Xu2021PoF}
T.~Xu, S.~Liao, Accurate predictions of chaotic motion of a free fall disk,
  Physics of Fluids 33 (2021) 037111.

\bibitem{li2021one}
X.~Li, X.~Li, S.~Liao, One family of 13315 stable periodic orbits of
  non-hierarchical unequal-mass triple systems, SCIENCE CHINA Physics,
  Mechanics \& Astronomy 64~(1) (2021) 1--6.

\bibitem{Liao2021arXiv}
S.~Liao, S.~Qin, Ultra-chaos: an insurmountable objective obstacle of
  reproducibility and replication, arXiv Preprint 2110.00143.

\bibitem{oyanarte1990mp}
P.~Oyanarte, {MP}-a multiple precision package, Computer Physics Communications
  59~(2) (1990) 345--358.

\bibitem{NewScientist2017}
L.~Crane, Infamous three-body problem has over a thousand new solutions, New
  Scientist(https://www.newscientist.com/article/2148074-infamous-three-body-problem-has-over-a-thousand-new-solutions/).

\bibitem{NewScientist2018}
C.~Whyte, Watch the weird new solutions to the baffling three-body problem, New
  Scientist(https://www.newscientist.com/article/2170161-watch-the-weird-new-solutions-to-the-baffling-three-body-problem/).

\bibitem{Stankevich2020Chaos}
N.~Stankevich, A.~Kazakov, S.~Gonchenko, Scenarios of hyperchaos occurrence in
  4{D} {R}\"{o}ssler system, Chaos 30 (2020) 123129.

\bibitem{chacon2008spatiotemporal}
R.~Chac{\'o}n, A.~Bellor{\'\i}n, L.~Guerrero, J.~Gonz{\'a}lez, Spatiotemporal
  chaos in sine-{G}ordon systems subjected to wave fields: Onset and
  suppression, Physical Review E 77~(4) (2008) 046212.

\bibitem{Keller1995Surveys}
J.~B. Keller, D.~W. Mclaughlin, G.~C. Papanicolaou, {Surveys in Applied
  Mathematics}, Plenum Press, 1995.

\bibitem{ferre2017localized}
M.~A. Ferr{\'e}, M.~G. Clerc, S.~Coulibally, R.~G. Rojas, M.~Tlidi, Localized
  structures and spatiotemporal chaos: comparison between the driven damped
  sine-{G}ordon and the {L}ugiato-{L}efever model, The European Physical
  Journal D 71~(6) (2017) 172.

\bibitem{Eiswirth1992CPL}
M.~Eiswirth, T.~M. Kruel, G.~Ertl, F.~W. Schneider, Hyperchaos in a chemical
  reaction, Chemical Physics Letters 193~(4) (1992) 305.

\bibitem{Kapitaniak1995CSF}
T.~Kapitaniak, K.~E. Thylwe, I.~Cohen, J.~Wjewoda, Chaos - hyperchaos
  transition, Chaos, Solitons \& Fractals 5~(10) (1995) 2003 -- 2011.

\bibitem{Baier1995PRE}
G.~Baier, S.~Sahle, Design of hyperchaotic flows, Physical Review E 51 (1995)
  2712 -- 2714.

\bibitem{Henriksen1994}
R.~N. Henriksen, Compressible turbulence or spatiotemporal chaos?, Astrophysics
  and Space Science 221 (1994) 25 -- 39.

\bibitem{Monya2016Nature}
M.~Baker, D.~Penny, Is there a reproducibility crisis?, Nature 533.

\bibitem{Colin2016Science}
F.~C. Camerer, {\em et al.}, Evaluating replicability of laboratory experiments
  in economics, Science 351~(6280) (2016) 1433.

\bibitem{Peng2011Science}
R.~D. Peng, Reproducible research in computational science, Science 334 (2011)
  1226.

\bibitem{Mesnard2017OSE}
O.~Mesnard, L.~A. Barba, Reproducible and replicable computational fluid
  dynamics: It is harder than you think, Computing in Science \& Engineering
  19~(4) (2017) 44 -- 55.

\bibitem{Benjamin2018NatureHB}
D.~J. Benjamin, {\em et al.}, Redefine statistical significance, Nature Human
  Behaviour 2 (2018) 6 -- 10.
\end{thebibliography}
\end{document}